\pdfoutput=1  % comment out if want postscript output
\documentclass[aps,prc,twocolumn,showpacs,nofootinbib]{revtex4}

\usepackage{xspace}
\usepackage{graphicx}
\usepackage{amsmath}
\usepackage{dcolumn}
\usepackage{longtable}
\usepackage{bm}
\usepackage{wasysym}
\usepackage[]{color}
\usepackage{textcomp}    % used for 1/4 symbol
\usepackage{hyperref}
\usepackage{txfonts}
\RequirePackage{ifpdf}

\newcommand{\nuc}[2]{${}^{#1}$#2}
\newcommand{\mnuc}[2]{{}^{#1}{\protect\text{#2}}}
\newcommand{\E}[1]{$\,\times$\,10$^{#1}$}
\renewcommand{\today}{\number\day\space\ifcase\month\or January
 \or February\or March\or April\or May\or June\or July\or August
 \or September\or October\or November\or December\fi\space\number\year}
\newcommand{\mc}[3]{\multicolumn{#1}{#2}{#3}}
\newcommand{\mco}[1]{\multicolumn{1}{c}{#1}}
\newcommand{\degreesC}{$\,^{\circ}$C\xspace}
\newcommand{\jour}[4]{{#1} {\bf #2}, {#3} ({#4})}

\begin{document}

\title{
       Measurement of the solar neutrino capture rate with gallium metal. \\
       III: Results for the 2002--2007 data-taking period.
      }

\author{
        J.\,N.\,Abdurashitov, V.\,N.\,Gavrin, V.\,V.\,Gorbachev,
        P.\,P.\,Gurkina, T.\,V.\,Ibragimova, A.\,V.\,Kalikhov,
        N.\,G.\,Khairnasov, T.\,V.\,Knodel, I.\,N.\,Mirmov, A.\,A.\,Shikhin,
        E.\,P.\,Veretenkin, V.\,E.\,Yants, and G.\,T.\,Zatsepin
       }
\affiliation{
             Institute for Nuclear Research,
             Russian Academy of Sciences, 117312 Moscow, Russia
            }

\author{
        T.\,J.\,Bowles, S.\,R.\,Elliott, and W.\,A.\,Teasdale
       }
\affiliation{
             Los Alamos National Laboratory, Los Alamos, New Mexico
             87545, USA
            }

\author{
        J.\,S.\,Nico
       }
\affiliation{
             National Institute of Standards and Technology, Stop
             8461, Gaithersburg, Maryland 20899, USA
            }

\author{
        B.\,T.\,Cleveland
       }
\thanks{
        Corresponding author.  Present address: SNOLAB, PO Box 159,
        Lively, Ontario \hspace{.3em} P3Y 1M3, Canada; bclevela@snolab.ca.
       }
\author{
        J.\,F.\,Wilkerson
       }
\affiliation{
             University of Washington, Seattle, Washington 98195, USA
            }

\author{(The SAGE Collaboration)}
\noaffiliation

%\date{\today}
\date{11 July 2009}

\begin{abstract}

The Russian-American experiment SAGE began to measure the solar
neutrino capture rate with a target of gallium metal in December~1989.
Measurements have continued with only a few brief interruptions since
that time.  In this article we present the experimental improvements
in SAGE since its last published data summary in December~2001.
Assuming the solar neutrino production rate was constant during the
period of data collection, combined analysis of 168 extractions
through December~2007 gives a capture rate of solar neutrinos with
energy more than 233~keV of $65.4 ^{+3.1}_{-3.0}$ (stat)
$^{+2.6}_{-2.8}$ (syst)~SNU.  The weighted average of the results of
all three Ga solar neutrino experiments, SAGE, Gallex, and GNO, is now
$66.1 \pm 3.1$~SNU, where statistical and systematic uncertainties
have been combined in quadrature.  During the recent period of data
collection a new test of SAGE was made with a reactor-produced
\nuc{37}{Ar} neutrino source.  The ratio of observed to calculated
rates in this experiment, combined with the measured rates in the
three prior \nuc{51}{Cr} neutrino-source experiments with Ga, is $0.87
\pm 0.05$.  A probable explanation for this low result is that the
cross section for neutrino capture by the two lowest-lying excited
states in \nuc{71}{Ge} has been overestimated.  If we assume these
cross sections are zero, then the standard solar model including
neutrino oscillations predicts a total capture rate in Ga in the range
of 63~SNU to 66~SNU with an uncertainty of about 4\%, in good
agreement with experiment.  We derive the current value of the
neutrino flux produced in the Sun by the proton-proton fusion
reaction to be $\phi^{\astrosun}_{pp}=(6.0 \pm 0.8) \times
10^{10}$/(cm$^2$~s), which agrees well with the $pp$ flux predicted by
the standard solar model.  Finally, we make several tests and show
that the data are consistent with the assumption that the solar
neutrino production rate is constant in time.

\end{abstract}


\maketitle

% Set page style
  \pagestyle{myheadings}
% Set text for header
  \markboth{\hfill SAGE results, Part III}
           {SAGE results, Part III \hfill}
% Set margin widths:
  \topmargin=-15mm % space from top of page to top of header + 25.4mm
% Set header position:
  \headheight=12pt
  \headsep=25pt         % space from bottom of header to top of text
% Set width of marginal paragraphs.
% Necessary if have two columns and paragraph is in left margin.
  \marginparwidth=35pt

\section{Introduction}

The SAGE experiment was built to measure the capture rate of solar
neutrinos by the reaction $\mnuc{71}{Ga} + \nu_e \rightarrow \,
\mnuc{71}{Ge} + e^-$ and thus to provide information to aid in
understanding the deficit of neutrinos observed in the \nuc{37}{Cl}
experiment \cite{clapj}, in which only about one-third of the solar
neutrino capture rate predicted by the standard solar model was
detected.  The feature that distinguishes the Ga experiment from all
other past or present solar neutrino detectors is its sensitivity to
the proton-proton fusion reaction, $p + p \rightarrow d + e^+ +
\nu_e$, which generates most of the Sun's energy.  Ga experiments have
provided the only direct measurement of the current rate of this
reaction.

A full description of the SAGE experiment and the results of each
measurement from its inception to December 1997 was presented in
Ref.~\cite{prc}.  Part II of this series, although not called by this
name, described the changes to the experiment and gave the results for
the period January 1998 to December 2001 \cite{jetp}.  In
Secs.~\ref{exp_procedures} and \ref{results} of the present article we
do the same for the six-year period January 2002 to December 2007.  We
then discuss the four neutrino source experiments with Ga in
Sec.~\ref{sources}, give the present interpretation of the SAGE
results in Sec.~\ref{interpretation}, derive the \textit{contemporary}
value of the neutrino flux produced by the proton-proton fusion
reaction in Sec.~\ref{pp_flux}, and present a brief consideration of
the question of possible time variation in the data in
Sec.~\ref{variation}.

In addition to SAGE, there also existed a second Ga solar neutrino
experiment called Gallex.  It contained 30~tons of gallium in a
solution of GaCl$_3$ and measured the solar neutrino capture rate from
1991 to 1997.  In 1998 this experiment was reconstituted under the
name of GNO and it took data until 2003.  We give the results of these
experiments and combine them with the SAGE data in Sec.~\ref{results}.

\section{Experimental procedures}
\label{exp_procedures}

\subsection{Overview}

The SAGE experiment is in a dedicated deep-underground laboratory
excavated into the side of Mt.\ Andyrchi in the northern Caucasus
mountains of Russia.  The rock overburden is equivalent to 4700~m of
water and the measured muon flux at the location of the experiment is
$(3.03 \pm 0.10) \times 10^{-9}/$(cm$^2$~s).

The mass of gallium used in SAGE at the present time is 50~tonnes.  It
is contained in seven chemical reactors which are heated to 30\degreesC so
the gallium metal remains molten.  A measurement of the solar neutrino
capture rate begins by adding to the gallium a stable Ge carrier.  The
carrier is a Ga-Ge alloy with a known Ge content of approximately
$350~\mu$g and is distributed equally among all reactors.  The reactor
contents are stirred thoroughly to disperse the Ge throughout the Ga
mass.  After a typical exposure interval of one month, the Ge carrier
and \nuc{71}{Ge} atoms produced by solar neutrinos and background
sources are chemically extracted from the Ga.  The final step of the
chemical procedure is the synthesis of germane (GeH$_4$), which is
used as a proportional counter fill gas with an admixture of
(90--95)\% Xe.  The total efficiency of extraction is the ratio of
mass of Ge in the germane to the mass of initial Ge carrier and is
typically $(95 \pm 3)\%$.

\subsection{Extraction of Ge from Ga}

The extraction procedures from 1990 to 1997 are described in
Ref.~\cite{prc}.  At the beginning of 1998 some minor modifications
were made as described in Ref.~\cite{ArPRC}.

Beginning with the December 2005 extraction, the carrier used to
measure the extraction efficiency was isotopically enriched in either
\nuc{72}{Ge} or \nuc{76}{Ge}.  At the end of each extraction a sample
was taken from the final extraction solution and this sample was
analyzed with an inductively-coupled plasma mass spectrometer to
determine the fractional content of the various Ge isotopes.  The
efficiency of Ge extraction from the Ga metal was then calculated
using the method outlined in Appendix \ref{ext_eff_calc}.  This
procedure for determining the extraction efficiency has the advantage
that it gives a direct measure of any Ge that may enter the sample
from unknown sources.

\subsection{\texorpdfstring{Counting of $^{\bf71}$G\lowercase{e}}%
                           {Counting of 71Ge}}

\nuc{71}{Ge} decays to \nuc{71}{Ga} by pure electron capture with a
half life of 11.4 days.  Two peaks are observed in the proportional
counter--the $K$~peak at 10.4~keV and the $L$~peak at 1.2~keV.  The
counter containing the GeH$_4$ from the extraction is placed in the
well of a NaI detector that is within a large passive shield and is
counted for a typical period of 6 months.  To reduce the influence of
\nuc{222}{Rn}, the volume inside the shield around the counters is
purged with boil-off gas from a dewar filled with liquid nitrogen.

A completely redesigned proportional counter \cite{Yants} began to be
used with the extraction of April 2001 and has been used for all but
two extractions since that time.  In contrast to the usual counters
with a solid cathode, the cathode of the new counters is made from
vapor-deposited carbon, thus eliminating the usual dead volume behind
the cathode.  The dead volume is further reduced and end effects are
nearly eliminated by curving inwards the regions of the counter where
the cathode ends.  The cathode and anode leads are sealed into the
counter body with Mo ribbon that makes the counter leak free and
ensures excellent gain stability.  The cathode is so thin that the
counter body is transparent, making it possible to visually inspect
all the internal counter parts.

During 2004--2005 an extensive series of measurements of the
efficiency of these new counters was made.  The methods of measurement
were described in Ref.~\cite{prc} and counter fillings of \nuc{69}{Ge},
\nuc{71}{Ge}, and \nuc{37}{Ar} were used.  The measured volume
efficiency of the new counters was 96\% with a spread in efficiency of
only $\pm1\%$ for all counters of this type.  This should be compared
with an average volume efficiency of 89\% for our original counter
design.  Further, the fraction of events that is degraded in energy
was found to be significantly less than in the old design.  These
decreases in degraded fraction combined with the increase in volume
efficiency lead to a quite dramatic increase in efficiency for these
new counters compared to the old type, approximately 25\% in the $K$
peak and 10\% in the $L$ peak.

Another innovation in the new counter design is that the Suprasil
counter body is etched in hydrofluoric acid to a thickness of
$\sim$0.2~mm.  This permits calibration of the counter with our
standard \nuc{55}{Fe} source over nearly its entire volume.  As an
undesired side effect, however, the thin body, combined with the very
thin cathode, makes these counters sensitive to low-energy x~rays from
local radioactivity.  To eliminate this response, a graded shield
consisting of an outer layer of 1~mm of Cu and an inner layer of 3~mm
of low-background acrylic (to absorb Cu x~rays) is placed over the
counter body during measurement with \nuc{71}{GeH$_4$}.

\begin{figure}
\centering

\ifpdf
 \includegraphics*[width=\hsize,viewport=26 18 220 275]{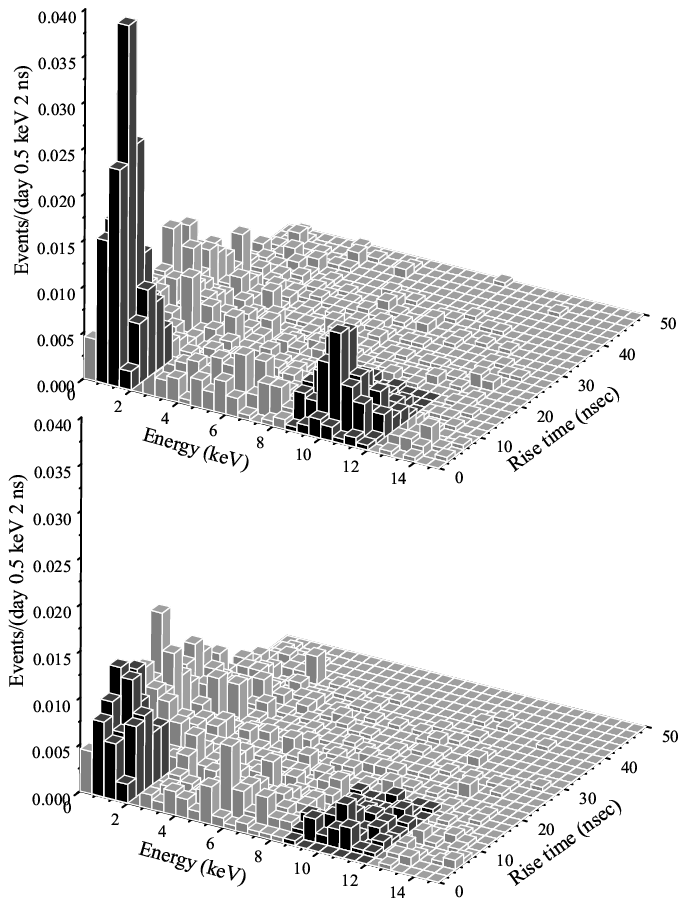}
\else
 \includegraphics*[width=\hsize,bb=26 18 220 275]{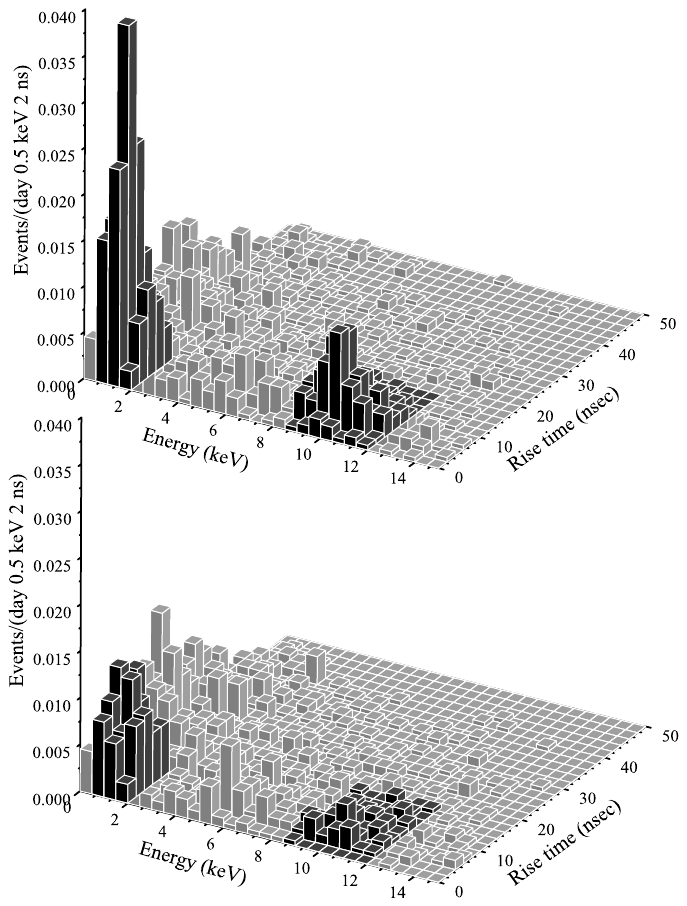}
\fi

\caption{Upper panel: Count rate vs energy and rise time for events
during the first 30 days of counting.  Regions where the $L$ and $K$
peaks are predicted to occur based on \nuc{55}{Fe} calibrations are
shown darkened.  There are 427 counts in the $L$-peak region and 287
counts in the $K$-peak region.  The counts in both regions are a
combination of events from \nuc{71}{Ge} decay and background.  Lower
panel: Equivalent graph for all events that occurred during an equal
live time interval beginning at day 100 after extraction.  There are
226 counts in the $L$-peak region and 94 counts in the $K$-peak
region.}

\label{3dhist}
\end{figure}

The pulses from the proportional counter are sent to a fast transient
analyzer where they are digitized for 800~ns after pulse onset at two
different gains, one chosen for the $L$~peak and the other for the $K$
peak.  The transient digitizer serves to differentiate fast-rising
\nuc{71}{Ge} pulses from generally slower-rising background pulses.
This can be seen by comparing the upper and lower panels of
Fig.~\ref{3dhist}, which show the pulses from the 77 extractions
that have been measured in the new proportional counters.  The upper
panel is for all events that pass the time cuts for Rn (see
Sec.~\ref{data_analysis}), are not high-voltage breakdown, do not
have a NaI coincidence, and occur during the first 30 days of
counting.  The total live time is 1999.8 days and there are 2063
events.  The lower panel of this figure shows the 1545 events that
occurred between days 100.0--130.1 (the same live time duration as in
the upper panel).  The fast-rising \nuc{71}{Ge} events in the $L$ and
$K$~peaks are evident in the upper panel but missing in the lower
panel because the \nuc{71}{Ge} has decayed away.

Aside from replacing some modules that failed, no changes were made to
the counting system electronics since their description in Ref.~\cite{prc}.

\begin{figure*}
\centering

\ifpdf
 \includegraphics*[width=0.8\hsize,viewport=30 38 442 214]{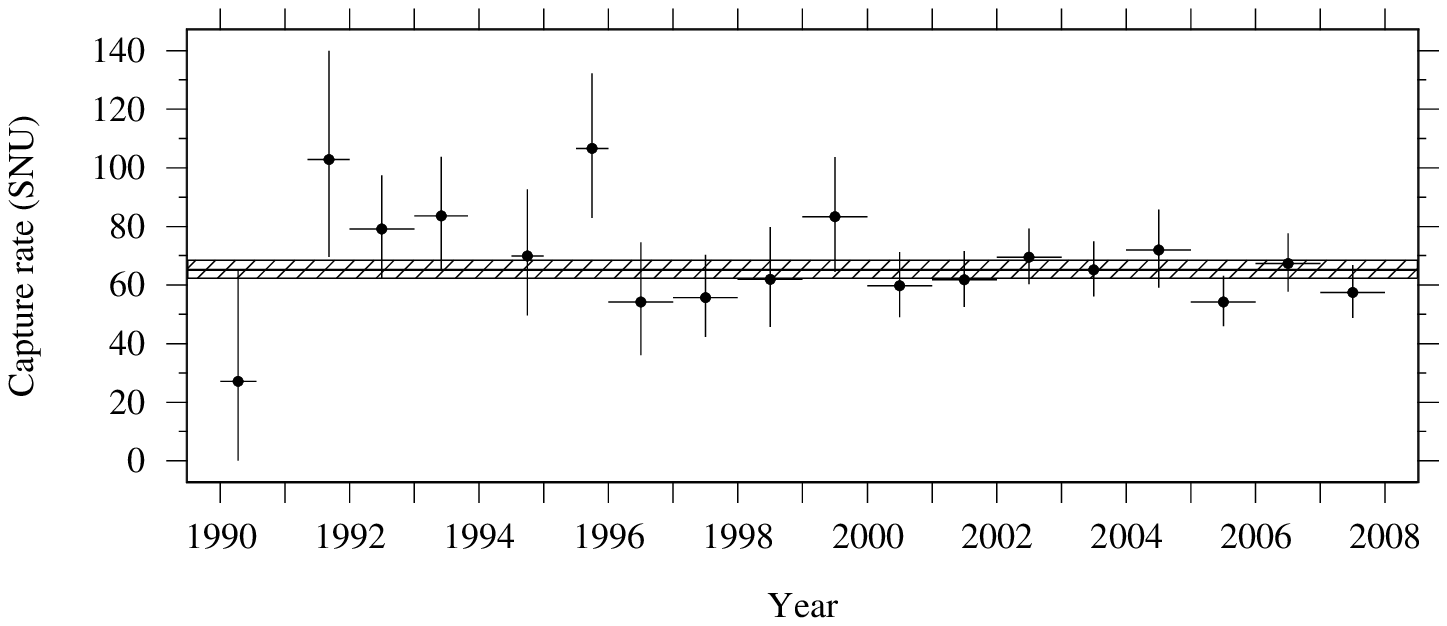}
\else
 \includegraphics*[width=0.8\hsize,bb=30 38 442 214]{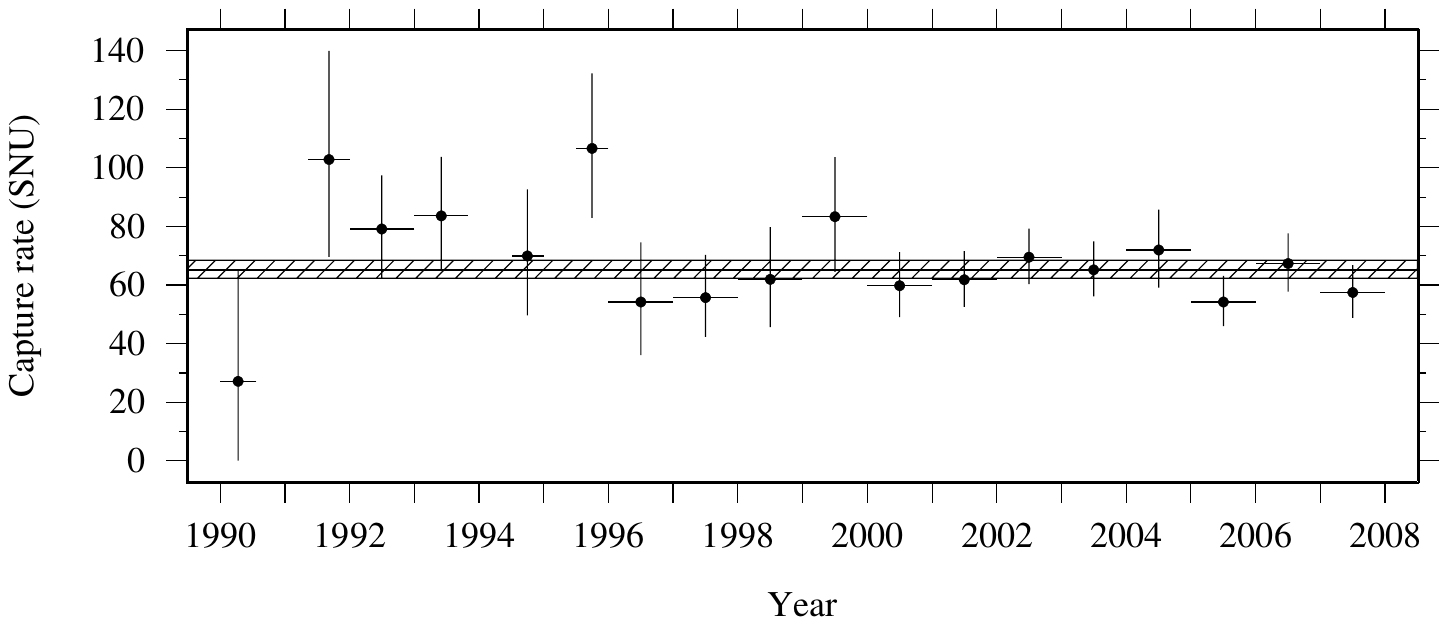}
\fi

\caption{Combined SAGE results for each year.  Shaded band is the
combined best fit and its uncertainty for all years.  Vertical error
bars are statistical with 68\% confidence.}

\label{yearly}
\end{figure*}

\subsection {Data analysis}
\label{data_analysis}

Based on criteria described in \cite{prc}, a group of events is
selected from each extraction that are candidate \nuc{71}{Ge} decays.
These events are fit to a maximum likelihood function \cite{max_like},
assuming that they originate from an unknown but constant-rate
background and the exponentially-decaying rate of \nuc{71}{Ge}.
Because only a few \nuc{71}{Ge} counts are detected from each
extraction, a single run result has a large statistical uncertainty
and thus little significance.

Several minor changes in the methods of analysis have occurred since
Part~II of this series.  These include the following:

\begin{itemize}

\item As discussed in Ref.~\cite{prc}, a small fraction of the decays of
\nuc{222}{Rn} are occasionally misidentified as pulses from
\nuc{71}{Ge}.  To reduce this effect for \nuc{222}{Rn} located
external to the counter we now delete all data that is acquired within
2.6~hours after counting begins.  In our initial analysis this time
cut was for only one hour.

\item To reduce the influence of \nuc{222}{Rn} that may enter the
counter when it is filled, we delete all data from 15~min before an
event that saturates the energy scale to 3~h after each saturated
event.  The SAGE measurements before September~1992, however, were
measured in counting systems that did not have the capability to
recognize saturated events \cite{prc}.  To reduce the number of these
false \nuc{71}{Ge} events produced by \nuc{222}{Rn} in these early
runs we determine for all subsequent runs the difference in capture
rate in the $K$~peak between the data analyzed with and without this
time cut and then subtract this difference from the result of each of
the runs before September~1992.

\item The predicted position and resolution of the $L$~peak from
calibrations was changed slightly from previous work.  These changes
were indicated by the results of a set of new calibrations made with
counters filled with \nuc{71}{Ge}.

\item The $L$- and $K$-peak shapes were changed from pure Gaussian to
Gaussian plus a degraded term.  The new functional form for the line
shape as a function of energy is
\begin{equation}
  \hspace{2em} % indents equation appropriately
  F(E) = h e^{-[(E-C)/(\sqrt{2}\sigma)]^2} + 
         hd\sqrt{\frac{\pi}{2}} \frac{\sigma}{C}
         \text{erfc} \left( \frac{E-C}{\sqrt{2}\sigma} \right),
\end{equation}
\noindent where $h,C,$ and $\sigma$ are the peak height, center, and
width and $d$ is a parameter related to the fraction of degraded
events.  The error function term here is the integral of the Gaussian
from energy $E$ to $\infty$; it is essentially flat below the peak,
monotonically decreases in the peak region, and is zero above the
peak.  This new line shape only makes a very small change to the
counting efficiency in the $L$~peak for a few runs whose energy window
width is obliged to be less than 2~full widths at half maximum.

\item For all runs after August~1992 the likelihood function was
modified to include a factor that weights each event according to its
measured energy.  This requires knowledge of the energy distribution
for \nuc{71}{Ge} pulses and for background events, both of which can
be determined from the long duration of counting data that we have
accumulated.  When this method is applied, it is found that the
overall statistical uncertainty decreases by (0.1-0.2)~SNU, but the
systematic uncertainty increases by $\sim$0.1~SNU.

\end{itemize}

These changes in analysis methods have been applied to all data.

\begin{table} \caption{Summary of known systematic effects and their
uncertainties.  SNU values for extraction and counting efficiency are
based on a rate of 65.4~SNU.}

\label{systematics}
\begin{tabular*}{\hsize}{@{} l @{\extracolsep{\fill}} r r}
\hline
\hline
                                        & \multicolumn{2}{c}{Uncertainty} \\ \cline{2-3}
Origin of uncertainty                   & \mc{1}{r}{in percent} & \mc{1}{r}{in SNU} \\
\hline
\multicolumn{3}{l}{Extraction efficiency}                               \\
 \hspace{2em} Ge carrier mass           & $\pm 2.1\%    $ & $\pm 1.4  $ \\
 \hspace{2em} Mass of extracted Ge      & $\pm 2.5\%    $ & $\pm 1.6  $ \\
 \hspace{2em} Residual Ge carrier       & $\pm 0.8\%    $ & $\pm 0.5  $ \\
 \hspace{2em} Ga mass                   & $\pm 0.3\%    $ & $\pm 0.2  $ \\
 \hspace{1em} Total (extraction)        & $\pm 3.4\%    $ & $\pm 2.2  $ \\
 \hline
\multicolumn{3}{l}{Counting efficiency}                                 \\
 \hspace{2em} Volume efficiency         & $\pm 1.0\%    $ & $\pm 0.7  $ \\
 \hspace{2em} End losses                & $\pm 0.5\%    $ & $\pm 0.3  $ \\
 \hspace{2em} Monte Carlo interpolation & $\pm 0.3\%    $ & $\pm 0.2  $ \\
 \hspace{2em} Shifts of gain            & $-1.1\%       $ & $+0.7     $ \\
 \hspace{2em} Resolution                & $+0.5\%,-0.7\%$ & $-0.3,+0.5$ \\
 \hspace{2em} Rise time limits          & $\pm 1.0\%    $ & $\pm 0.7  $ \\
 \hspace{2em} Lead and exposure times   & $\pm 0.8\%    $ & $\pm 0.5  $ \\
 \hspace{1em} Total (counting)          & $+1.8\%,-2.1\%$ & $-1.2,+1.4$ \\
 \hline
\multicolumn{3}{l}{Nonsolar neutrino production of \nuc{71}{Ge}} \rule{0pt}{2.5ex} \\
 \hspace{2em} Fast neutrons             &                 & $<- 0.02  $ \\
 \hspace{2em} \nuc{232}{Th}             &                 & $<- 0.04  $ \\
 \hspace{2em} \nuc{226}{Ra}             &                 & $<- 0.7   $ \\
 \hspace{2em} Cosmic-ray muons          &                 & $<- 0.7   $ \\
 \hspace{1em} Total (nonsolar)          &                 & $<- 1.0   $ \\
 \hline
\multicolumn{3}{l}{Background events that mimic \nuc{71}{Ge}} \rule{0pt}{2.5ex}    \\
 \hspace{2em} Internal \nuc{222}{Rn}    &                 & $<- 0.2   $ \\
 \hspace{2em} External \nuc{222}{Rn}    &                 & $   0.0   $ \\
 \hspace{2em} Internal \nuc{69}{Ge}     &                 & $<- 0.6   $ \\
 \hspace{1em} Total (background events) &                 & $<- 0.6   $ \\
 \hline
Energy weighting in analysis            &                 & $\pm0.1   $ \\
 \hline
 Total                                  &                 & $-2.8,+2.6$ \\
\hline
\hline
\end{tabular*}
\end{table}

\section{Results}
\label{results}

\nuc{71}{Ge} has been extracted from the Ga target to measure the
solar neutrino capture rate every month from January 2000 to the
present time.  We even were able to make six solar extractions during
the time of the \nuc{37}{Ar} neutrino source experiment in 2004 by
sending the samples to Gran Sasso.  In a cooperative effort
\cite{BNOLNGS}, the GNO collaboration synthesized GeH$_4$ and measured
the samples in their counting system.

% In contrast to previous publications, we choose to not present here
% the usual tables of extraction data, counting data, and results.  The
% reader interested in this information can obtain it from the SAGE web
% site
% \href{http://ewi.npl.washington.edu/SAGE/sage.html}{http://ewi.npl.washington.edu/SAGE/sage.html}.

The results for each individual extraction are tabulated in
Appendix~\ref{results_table} and the combined result of each year of
SAGE data since its beginning is shown in Fig.~\ref{yearly}.

The systematic uncertainties in the experiment have been considered in
detail in Ref.~\cite{prc,jetp} and the most recent values are given in
Table~\ref{systematics}.  The only significant changes from our
previous articles are due to the new proportional counters.  Their
high stability and efficiency have led to a considerable reduction of
the uncertainties associated with counting.

In radiochemical experiments the capture rate has been conventionally
expressed in ``SNU units'', defined as one neutrino capture per second
in a target that contains $10^{36}$ atoms of the neutrino-absorbing
isotope, in our case \nuc{71}{Ga}.  For all SAGE data from January
1990 through December 2007 (168~runs and 310 separate counting sets)
the global best fit capture rate is $65.4 ^{+3.1}_{-3.0}$~SNU, where
the uncertainty is statistical only.  If one considers the $L$-peak
and $K$-peak data separately, the results are $67.2
^{+4.8}_{-4.6}$~SNU and $64.0 ^{+4.1}_{-4.0}$~SNU, respectively.  The
agreement between the two peaks serves as a strong check on the
robustness of the event selection criteria.  Including the systematic
uncertainty, our overall result is $65.4 ^{+3.1}_{-3.0} \text{~(stat)}
^{+2.6}_{-2.8} \text{~(syst)}$~SNU.

As further evidence that we are truly counting \nuc{71}{Ge}, we can
allow the decay constant during counting to be a free variable in the
maximum likelihood fit, along with the combined \nuc{71}{Ge}
production rate and all the background rates.  The best fit half-life
to all selected events in both $L$ and $K$ peaks is then $11.5 \pm
0.9$ (stat) days, in agreement with the measured value \cite{HAM85}
of $11.43 \pm 0.03$~days.

The waveform data from the Gallex experiment has recently been
re-evaluated by Kaether using a new pulse-shape analysis method
\cite{Kaether_thesis} and the result is $73.1 ^{+6.1 +3.7}_{-6.0
-4.1}$~SNU.  The result of the GNO experiment was $62.9 ^{+5.5
+2.5}_{-5.3 -2.5}$~SNU \cite{gno_final}.  If we combine the
statistical and systematic uncertainties in quadrature, then the
weighted combination of all the Ga experiments, SAGE, Gallex, and GNO,
is
\begin{equation}
66.1 \pm 3.1 \text{~SNU. \quad \quad (Present Ga experiment result.)}
\label{Ga_result}
\end{equation}

\section{Source experiments}
\label{sources}

\begin{figure}
\centering

\ifpdf
 \includegraphics[width=\hsize,viewport=21 31 447 280]{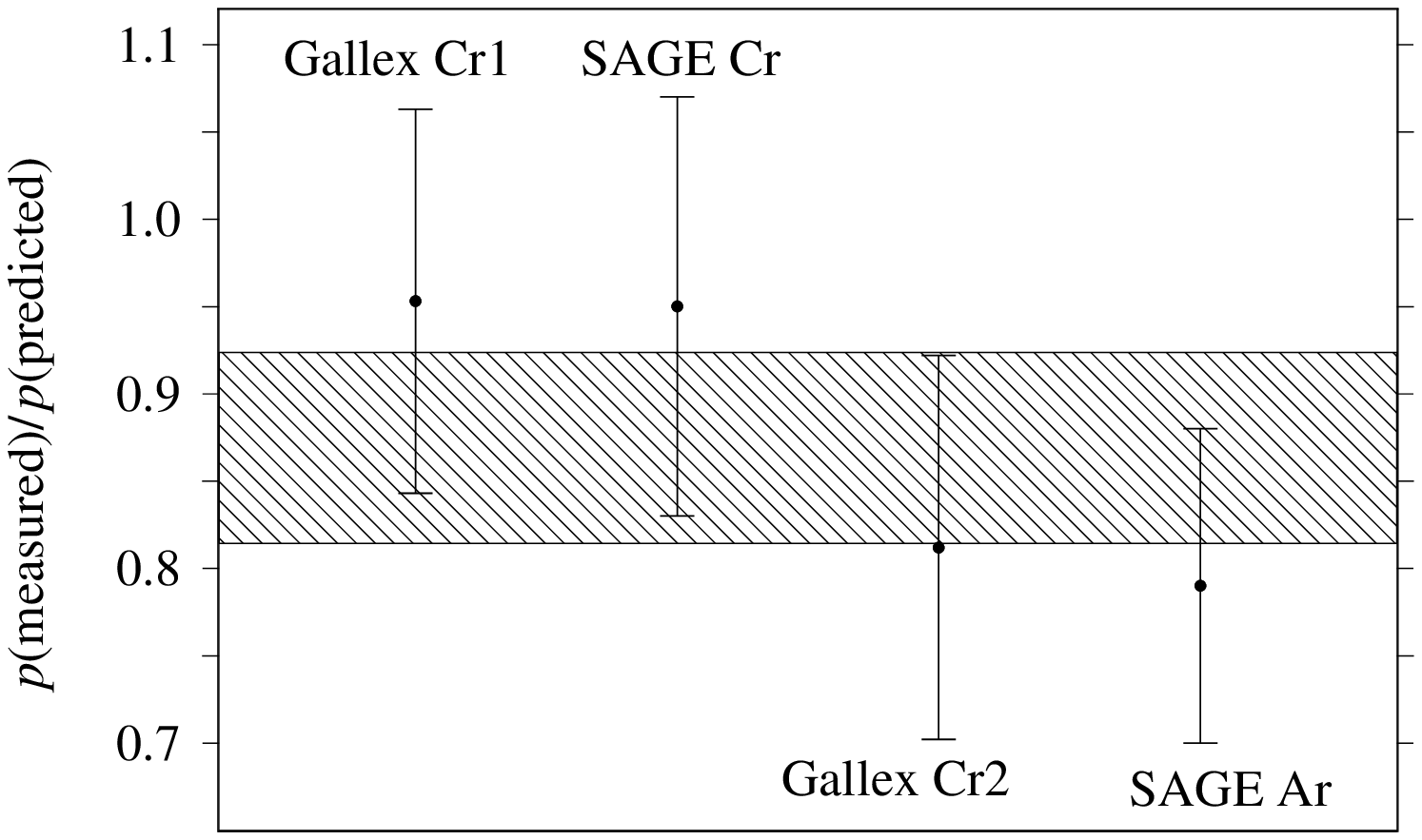}
\else
 \includegraphics[width=\hsize,bb=21 31 447 280]{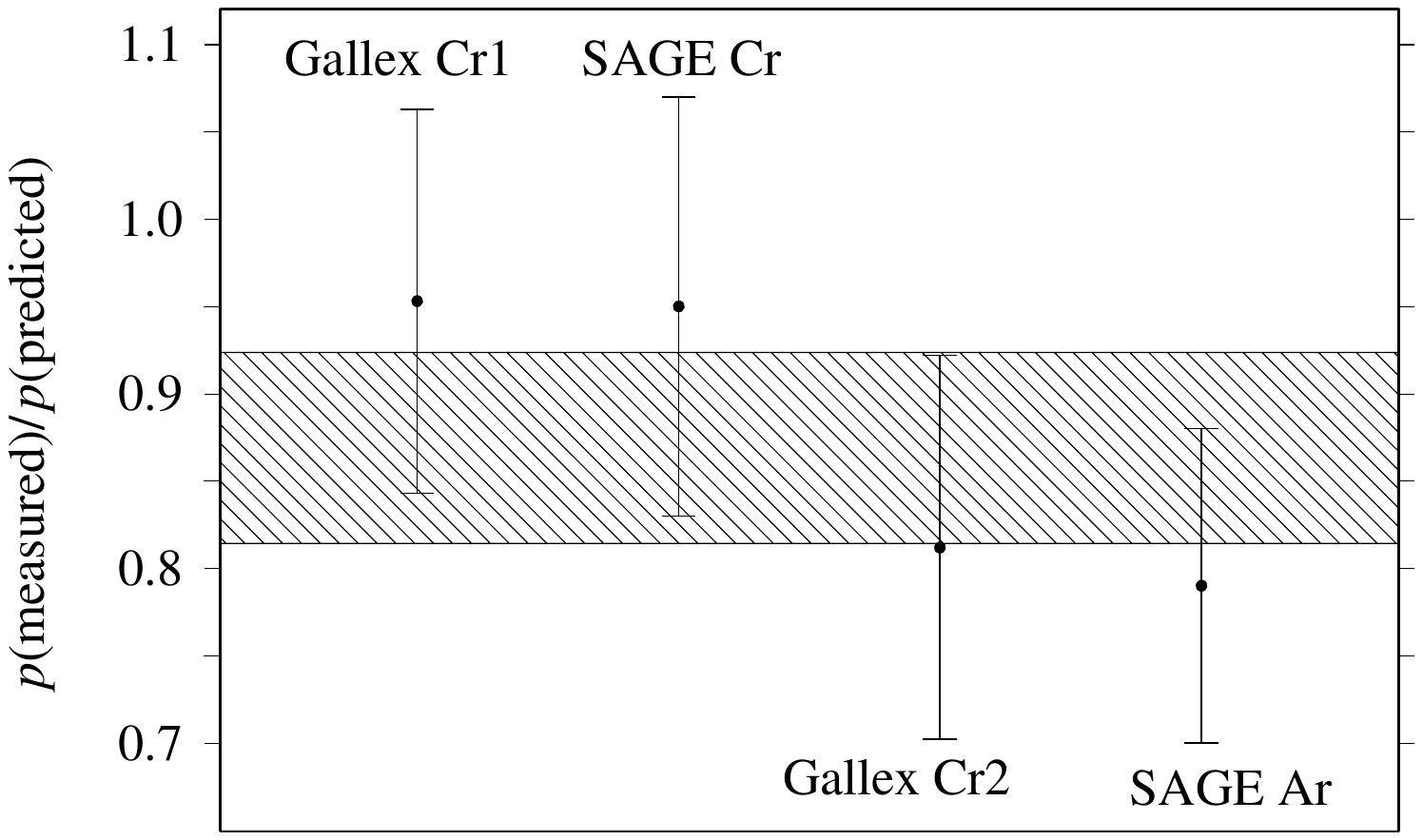}
\fi

\caption{Results of all neutrino source experiments with Ga.  Gallex
results are from the recent pulse shape analysis of Kaether {\protect
\cite{Kaether_thesis}}; SAGE results are from Refs.~{\protect \cite{CrPRC}}
and {\protect \cite{ArPRC}}.  Hashed region is the weighted average of
the four experiments.}

\label{src_res}
\end{figure}

The experimental procedures of the SAGE and Gallex experiments,
including the chemical extraction, counting, and analysis techniques,
have been checked by exposing the gallium target to reactor-produced
neutrino sources whose activity was close to 1~MCi.  SAGE has
irradiated about 25\% of their target with a \nuc{51}{Cr} source
\cite{CrPRC} and an \nuc{37}{Ar} source
\cite{Haxton,Gavrin_preprint,ArPRC} and Gallex has twice used
\nuc{51}{Cr} sources to irradiate their entire target
\cite{gallex_cr51}.  The results, expressed as the ratio $R$ of the
measured \nuc{71}{Ge} production rate to that expected due to the
source strength, are shown in Fig.~\ref{src_res}.  The weighted
average value of the ratio for the four experiments is $R = 0.87 \pm
0.05$, more than two standard deviations less than unity.  Although
the distribution of results is somewhat unusual, with none of the
central values from the four measurements lying within the $1\sigma$
band around the weighted average, the quality of fit to the average
value is quite high ($\chi^2/\text{DOF}=1.9/3$, GOF = 59\%).

We can suggest several possibilities for the unexpectedly low result
in the source experiments:

\begin{enumerate}

\item We do not correctly know the various efficiency factors that
enter into the calculation of the production rate, namely the
extraction efficiency and the counting efficiency.  Both SAGE and
Gallex have, however, made many ancillary experiments
\cite{jetp,gallex_cr51} that have established with high probability
that these efficiencies and their accompanying systematic
uncertainties are well determined.  These tests have also proven that
there are no substantial errors in the methods used to select
\nuc{71}{Ge} events or in the methods of analysis.  Further, the
\nuc{71}{As} experiment of Gallex \cite{gallex71as} has ruled out any
``hot-atom'' chemical effects that might make the \nuc{71}{Ge} atoms
produced by neutrino capture difficult to extract.  We thus very
strongly doubt that the low average result of the source experiments
is due to incorrect knowledge of efficiencies, errors in event
selection, improper functioning of the counting systems, or errors in
analysis.

\item A statistical fluctuation.  A $\chi^2$ test of the compatability
of the four source experiments to $R=1.0$ gives
$\chi^2/\text{DOF}=7.7/3$, whose probability is 5.3\%.  The probability
is small, but still quite possible.

\item Electron neutrinos disappear due to a real physical effect of
unknown origin.  Some possibilities that have been suggested are a
transition to sterile neutrinos \cite{Giunti} or quantum decoherence
in neutrino oscillations \cite{Farzan}.  

\item The production rate from the source is not as great as has been
assumed.  It is our opinion that this is the most likely cause of the
apparently low result in the source experiments.  As suggested by
Haxton \cite{Haxtoncross} it is quite possible that the cross sections
for neutrino capture to the two lowest excited states in \nuc{71}{Ge},
both of which can be reached using either \nuc{51}{Cr} or \nuc{37}{Ar}
sources, have been overestimated.  95\% of the capture rate with these
sources arises from the \nuc{71}{Ga} to \nuc{71}{Ge} ground-state
transition with 5\% due to transitions to the two excited states.
If the contribution of the excited states to the predicted rate were
to be zero then $R=p(\text{measured})/p(\text{predicted}) = 0.92 \pm
0.06$ and the fit to the expected value of 1.0 becomes quite
reasonable ($\chi^2/\text{DOF} = 4.58/3$, GOF = 21\%).

% R = (0.8689 +/- 0.0547)/0.95 = 0.9146 +/- 0.0575

\end{enumerate}

A concern in this context is the implication of the apparently low
result of the source experiments on the solar neutrino result given in
Eq.~(\ref{Ga_result}).  It is difficult to address this concern
because we do not understand why the source experiments give a lower
result than expected.  If we suppose that the cause is item 1 in the
list above, then the rate in Eq.~(\ref{Ga_result}) should be divided
by the factor 0.87, i.e., we should add 15\% to the systematic
uncertainty.  But, as stated above, we consider that explanation for
the apparent discrepancy in the source experiments to be very
unlikely.  However, if we suppose that the cause of the low
result in the source experiments is any of the other items in the list
above, then the source experiments have no bearing on the solar
neutrino result and the rate in Eq.~(\ref{Ga_result}) should not be
changed.  Because we do not know why the source experiments appear to
be low, we can only caution the reader to accept the result in
Eq.~(\ref{Ga_result}) on a provisional basis, subject to the caveats
that not all effects in the emission of neutrinos from the Sun and the
capture of neutrinos by \nuc{71}{Ga} may be fully understood.

Based on the information given in the definitive article of Bahcall
\cite{gacross} on the neutrino capture cross section of \nuc{71}{Ga},
we have approximately calculated the cross section if we assume zero
strength for capture to the first two excited states of \nuc{71}{Ge}.
These results are given in Appendix~\ref{sigma_app} and will be used
as a working hypothesis in what follows.

\begin{table}

\caption{Solar neutrino fluxes calculated by the standard solar model
of Bahcall, Pe\~na-Garay, and Serenelli {\protect \cite{PGS09}} for
two different conservative choices of heavy element composition,
labeled GS98 {\protect \cite{GS98}} (high metallicity) and AGS05
{\protect \cite{AGS05}} (low metallicity).  The spectrum components
refer to the nuclear reaction from which they originate.  Units of
flux are $10^{10}\,\,(pp)$, $10^{9}$ (\nuc{7}{Be}), $10^{8}\,\,(pep$,
\nuc{13}{N}, \nuc{15}{O}), $10^{6}$ (\nuc{8}{B}, \nuc{17}{F}), and
$10^{3}\,\,(hep)$ cm$^{-2}$s$^{-1}$.  The uncertainty values are at
68\% confidence.}

\def\arraystretch{1.2}
\label{solar_flux}
\centering
\begin{tabular*}{1.00\hsize}{@{} l @{\extracolsep{\fill}} c c @{}}
\hline
\hline
Spectrum      & \mc{2}{c}{Flux $\phi^{\astrosun}_i$}                             \\ \cline{2-3}
component $i$ & GS98                            & AGS05                          \\
\hline
$pp$          & $ 5.97 (1^{+0.006}_{-0.006}) $  & $ 6.04 (1^{+0.005}_{-0.005}) $ \\
$pep$         & $ 1.41 (1^{+0.011}_{-0.011}) $  & $ 1.45 (1^{+0.010}_{-0.010}) $ \\
\nuc{7}{Be}   & $ 5.07 (1^{+0.06 }_{-0.06 }) $  & $ 4.55 (1^{+0.06 }_{-0.06 }) $ \\
\nuc{13}{N}   & $ 2.88 (1^{+0.15 }_{-0.15 }) $  & $ 1.89 (1^{+0.14 }_{-0.13 }) $ \\
\nuc{15}{O}   & $ 2.15 (1^{+0.17 }_{-0.16 }) $  & $ 1.34 (1^{+0.16 }_{-0.15 }) $ \\
\nuc{17}{F}   & $ 5.82 (1^{+0.19 }_{-0.17 }) $  & $ 3.25 (1^{+0.16 }_{-0.15 }) $ \\
\nuc{8}{B}    & $ 5.94 (1^{+0.11 }_{-0.11 }) $  & $ 4.72 (1^{+0.11 }_{-0.11 }) $ \\
$hep$         & $ 7.90 (1^{+0.15 }_{-0.15 }) $  & $ 8.22 (1^{+0.15 }_{-0.15 }) $ \\
\hline
\end{tabular*}
\end{table}

\begin{table*}
% Programs and calculations are in \solarnu\btc\cross4.

\caption{Factors needed to compute the solar neutrino capture rate in
\nuc{71}{Ga} and \nuc{37}{Cl} solar neutrino experiments.  The
uncertainty values are at 68\% confidence.  The parameter
$\langle\sigma_i^{\earth}\rangle$ is defined in
Eq.~\protect{(\ref{sigma_earth})}.}

\label{sigma_table}
\scriptsize{
\begin{tabular*}{1.00\hsize}{@{} l @{\extracolsep{\fill}} c c c c c c d c c c c c @{}}
\hline
\hline
     & Spect. &                           & \mc{3}{c}{Percent uncertainty in $\langle P^{ee}_i\rangle$ due to} & Total unc.                        & \mco{$\langle\sigma_i^{\earth}\rangle$} & \mc{4}{c}{Percent uncertainty in $\langle\sigma_i^{\earth}\rangle$ due to} & Total unc.                                \rule[2.5ex]{0pt}{0pt} \rule[-1ex]{0pt}{0pt} \\ \cline{4-6} \cline{9-12}
Exp. & comp.  & $\langle P^{ee}_i\rangle$ & $\Delta \text{m}^2_{12}$ & $\theta_{12}$ & $\theta_{13}$           & in $\langle P^{ee}_i\rangle$ (\%) & \mco{($10^{-46}\text{ cm}^2)$}          & $\sigma$    & $\Delta \text{m}^2_{12}$ & $\theta_{12}$ & $\theta_{13}$     & in $\langle\sigma_i^{\earth}\rangle$ (\%) \rule[2.5ex]{0pt}{0pt} \rule[-1ex]{0pt}{0pt} \\
\hline
\nuc{71}{Ga} & $pp$        & 0.561 & + 0.0,- 0.0 & + 2.2,- 2.8 & + 2.0,- 3.1 & + 3.0,- 4.2 &    11.75  & + 2.4,- 2.3 & + 0.0,- 0.0 & + 0.0,- 0.0 & + 0.0,- 0.0 & + 2.4,- 2.3 \rule[2.5ex]{0pt}{0pt} \\
             & $pep$       & 0.521 & + 0.3,- 0.3 & + 1.9,- 2.3 & + 1.9,- 2.9 & + 2.7,- 3.7 &   194.4   & +17  ,- 2.4 & + 0.0,- 0.0 & + 0.0,- 0.0 & + 0.0,- 0.0 & +17  ,- 2.4 \\
             & \nuc{7}{Be} & 0.542 & + 0.1,- 0.1 & + 2.1,- 2.6 & + 2.0,- 3.0 & + 2.9,- 4.0 &    68.21  & + 7.0,- 2.3 & + 0.0,- 0.0 & + 0.0,- 0.0 & + 0.0,- 0.0 & + 7.0,- 2.3 \\
             & \nuc{13}{N} & 0.545 & + 0.1,- 0.1 & + 2.1,- 2.6 & + 2.0,- 3.0 & + 2.9,- 4.0 &    56.83  & + 9.8,- 2.3 & + 0.0,- 0.0 & + 0.1,- 0.0 & + 0.0,- 0.0 & + 9.8,- 2.3 \\
             & \nuc{15}{O} & 0.535 & + 0.2,- 0.2 & + 2.0,- 2.5 & + 1.9,- 3.0 & + 2.8,- 3.9 &   107.1   & +13  ,- 2.3 & + 0.0,- 0.0 & + 0.1,- 0.1 & + 0.0,- 0.0 & +13  ,- 2.3 \\
             & \nuc{17}{F} & 0.535 & + 0.2,- 0.2 & + 2.0,- 2.5 & + 1.9,- 3.0 & + 2.8,- 3.9 &   107.7   & +13  ,- 2.3 & + 0.0,- 0.0 & + 0.1,- 0.1 & + 0.0,- 0.0 & +13  ,- 2.3 \\
             & \nuc{8}{B}  & 0.365 & + 0.7,- 0.6 & + 3.2,- 2.2 & + 1.8,- 2.7 & + 3.8,- 3.5 & 21400     & +32  ,-14   & + 0.2,- 0.2 & + 2.1,- 1.7 & + 0.1,- 0.1 & +32  ,-15   \\
             & $hep$       & 0.337 & + 0.5,- 0.5 & + 4.8,- 3.4 & + 1.8,- 2.8 & + 5.2,- 4.4 & 66000     & +33  ,-15   & + 0.2,- 0.2 & + 1.4,- 1.1 & + 0.1,- 0.1 & +33  ,-16   \\
\hline
\nuc{37}{Cl} & $pep$       & 0.521 & + 0.3,- 0.3 & + 1.9,- 2.3 & + 1.9,- 2.9 & + 2.7,- 3.7 &    16.00  & + 2.0,- 2.0 & + 0.0,- 0.0 & + 0.0,- 0.0 & + 0.0,- 0.0 & + 2.0,- 2.0 \rule[2.5ex]{0pt}{0pt} \\
             & \nuc{7}{Be} & 0.542 & + 0.1,- 0.1 & + 2.1,- 2.6 & + 2.0,- 3.0 & + 2.9,- 4.0 &     2.397 & + 2.0,- 2.0 & + 0.0,- 0.0 & + 0.0,- 0.0 & + 0.0,- 0.0 & + 2.0,- 2.0 \\
             & \nuc{13}{N} & 0.545 & + 0.1,- 0.1 & + 2.1,- 2.6 & + 2.0,- 3.0 & + 2.9,- 4.0 &     1.686 & + 2.0,- 2.0 & + 0.1,- 0.1 & + 0.1,- 0.1 & + 0.0,- 0.0 & + 2.0,- 2.0 \\
             & \nuc{15}{O} & 0.535 & + 0.2,- 0.2 & + 2.0,- 2.5 & + 1.9,- 3.0 & + 2.8,- 3.9 &     6.662 & + 2.0,- 2.0 & + 0.1,- 0.1 & + 0.2,- 0.1 & + 0.0,- 0.0 & + 2.0,- 2.0 \\
             & \nuc{17}{F} & 0.535 & + 0.2,- 0.2 & + 2.0,- 2.5 & + 1.9,- 3.0 & + 2.8,- 3.9 &     6.710 & + 2.0,- 2.0 & + 0.1,- 0.1 & + 0.2,- 0.1 & + 0.0,- 0.0 & + 2.0,- 2.0 \\
             & \nuc{8}{B}  & 0.365 & + 0.7,- 0.6 & + 3.2,- 2.2 & + 1.8,- 2.7 & + 3.8,- 3.5 & 10140     & + 3.7,- 3.7 & + 0.2,- 0.2 & + 2.4,- 1.9 & + 0.1,- 0.1 & + 4.4,- 4.1 \\
             & $hep$       & 0.337 & + 0.5,- 0.5 & + 4.8,- 3.4 & + 1.8,- 2.8 & + 5.2,- 4.4 & 40910     & + 3.7,- 3.7 & + 0.2,- 0.2 & + 1.5,- 1.2 & + 0.1,- 0.1 & + 4.0,- 3.9 \\
\hline
\hline
\end{tabular*}
} % end scriptsize
\end{table*}

\section{Interpretation of results}
\label{interpretation}

In contrast to all other present or past solar neutrino experiments,
the radiochemical Ga experiment, because of its low threshold of
233~keV, is sensitive to all components of the solar spectrum, from
the low-energy $pp$ neutrinos to the high-energy neutrinos produced in
the decay of \nuc{8}{B}.  In Table~\ref{solar_flux} we give the flux
of the various solar neutrino components at their production regions
in the Sun as calculated by Bahcall and collaborators
\cite{BSB06,PGS09}.  In this section we will estimate the neutrino
capture rate from each flux component and compare their total to the
measured rate.

The total capture rate $R$ of solar neutrinos in a radiochemical
experiment such as Ga is given by
\begin{equation} 
R = \int_{E_{\text{threshold}}}^{\infty} \sigma(E) \Phi^{\earth}(E)dE,
\label{rate_definition}
\end{equation}
\noindent where $\sigma(E)$ is the cross section of the
neutrino-capture reaction and $\Phi^{\earth}(E)$ is the total flux of
electron neutrinos at the Earth, which can be expressed as
\begin{equation}
\Phi^{\earth}(E) = \sum_i \phi^{\earth}_{i} S^{\earth}_i(E).
\label{flux_definition}
\end{equation}
\noindent In this expression the index $i$ refers to the various
nuclear reactions in the Sun that produce neutrinos ($pp$,
\nuc{7}{Be}, $pep$, \nuc{13}{N}, \nuc{15}{O}, \nuc{17}{F}, \nuc{8}{B},
and $hep$), $\phi^{\earth}_i$ is the amplitude of flux component $i$
at the Earth, and $S^{\earth}_i(E)$ is the spectrum of the $i$th
neutrino component at the Earth, each of which is normalized such that
$\int_0^{\infty}S^{\earth}_i(E)dE = 1$.  The neutrino spectrum at the
Earth is related to the spectrum produced in the Sun
$S^{\astrosun}_i(E)$ by
\begin{equation}
S^{\earth}_i(E) = A_i S^{\astrosun}_i(E) P^{ee}_i(E),
\label{spectrum}
\end{equation}
\noindent where $A_i$ is a constant of normalization and $P^{ee}_i(E)$
is the probability that an electron neutrino produced in the Sun by
reaction $i$ with energy $E$ will reach the Earth without a change of
flavor, commonly called the survival factor.  The physical origin for
the reduction of the electron component of the solar neutrino flux is
the now well-established mechanism of MSW neutrino oscillations
\cite{MSW}.  $P^{ee}_i(E)$ is different for each flux component as the
neutrinos are produced at different locations in the Sun and thus pass
through regions of different electron density during their travel to
the Earth.  $P^{ee}_i(E)$ can only be calculated if one knows where in
the Sun the neutrinos are made and thus requires a solar model.

We integrate Eq.~(\ref{spectrum}) and obtain $A_i = 1/\langle
P^{ee}_i\rangle$ where
\begin{equation}
\langle P^{ee}_i\rangle = \int_0^{\infty} S^{\astrosun}_i(E) P^{ee}_i(E) dE
\label{pee_average}
\end{equation}
\noindent is the spectrum-weighted average value of $P^{ee}_i$.  The
physical interpretation of $\langle P^{ee}_i\rangle$ is as the ratio
of the solar neutrino amplitudes at the surface of the Earth and at
the production point in the Sun:
\begin{equation}
\langle P^{ee}_i\rangle = \frac{\phi^{\earth}_i}{\phi^{\astrosun}_i}.
\label{flux_ratio}
\end{equation}

Combining these equations, and expressing $R$ as the sum of its
spectral components, $R = \sum_i R_i$, we have
\begin{equation}
R_i = \phi^{\astrosun}_i \langle \sigma_i^{\astrosun} \rangle,
\label{rate_equation}
\end{equation}
\noindent where
\begin{equation}
\langle \sigma_i^{\astrosun} \rangle =
   \int_{E_{\text{threshold}}}^{\infty} \sigma(E) S^{\astrosun}_i(E) P^{ee}_i(E)dE,
\label{sigma_solar}
\end{equation}
\noindent or, equivalently, if it is the flux at the Earth that is
assumed known,
\begin{equation}
R_i = \phi^{\earth}_i \langle \sigma_i^{\earth} \rangle,
\end{equation}
\noindent where
\begin{equation}
\langle \sigma_i^{\earth} \rangle = 
   \frac{\langle \sigma_i^{\astrosun} \rangle}{\langle P^{ee}_i\rangle}.
\label{sigma_earth}
\end{equation}

\begin{table*}
% This version is with PGS09 fluxes.
% Programs and calculations are in \solarnu\btc\caprate2.

\caption{Capture rates $R_i$ for Ga experiments calculated with fluxes
from Ref.\ {\protect \cite{PGS09}}.}

\label{rates}
\scriptsize{
\begin{tabular*}{1.00\hsize}{@{} l @{\extracolsep{\fill}} d c c c c c c c d c c c c c c @{}}
\hline
\hline
            & \mc{7}{c}{With GS98 composition}                                                                                      & & \mc{7}{c}{With AGS05 composition}                                                                                                                                 \\ \cline{2-8} \cline{10-16}
Spect.      & \mco{Cap.\ rate} & \mc{5}{c}{Percent uncertainty in rate due to}                                        & Total unc.  & & \mco{Cap.\ rate} & \mc{5}{c}{Percent uncertainty in rate due to}                                      & Total unc.   \rule[2.5ex]{0pt}{0pt} \rule[-1ex]{0pt}{0pt} \\ \cline{3-7} \cline{11-15}
comp.       & \mco{(SNU)}      & $\phi$      & $\sigma$    & $\Delta \text{m}^2_{12}$ & $\theta_{12}$ & $\theta_{13}$ & in rate (\%)& & \mco{(SNU)}      & $\phi$    & $\sigma$    & $\Delta \text{m}^2_{12}$ & $\theta_{12}$ & $\theta_{13}$ & in rate (\%) \rule[2.5ex]{0pt}{0pt} \rule[-1ex]{0pt}{0pt} \\
\hline
$pp$        & 39.35 & + 0.6,- 0.6 & + 2.4,- 2.3 & + 0.0,- 0.0 & + 2.2,- 2.8 & + 2.0,- 3.1 & + 3.9,- 4.8 & & 39.81 & + 0.5,- 0.5 & + 2.4,- 2.3 & + 0.0,- 0.0 & + 2.2,- 2.8 & + 2.0,- 3.1 & + 3.9,- 4.8 \\
$pep$       &  1.43 & + 1.1,- 1.1 & +17.0,- 2.4 & + 0.3,- 0.3 & + 1.9,- 2.3 & + 1.9,- 2.9 & +17.2,- 4.6 & &  1.47 & + 1.0,- 1.0 & +17.0,- 2.4 & + 0.3,- 0.3 & + 1.9,- 2.3 & + 1.9,- 2.9 & +17.2,- 4.5 \\
\nuc{7}{Be} & 18.73 & + 6.0,- 6.0 & + 7.0,- 2.3 & + 0.1,- 0.1 & + 2.1,- 2.6 & + 2.0,- 3.0 & + 9.7,- 7.5 & & 16.81 & + 6.0,- 6.0 & + 7.0,- 2.3 & + 0.1,- 0.1 & + 2.1,- 2.6 & + 2.0,- 3.0 & + 9.7,- 7.5 \\
\nuc{13}{N} &  0.89 & +15.0,-15.0 & + 9.8,- 2.3 & + 0.1,- 0.1 & + 2.1,- 2.6 & + 2.0,- 3.0 & +18.1,-15.7 & &  0.58 & +14.0,-13.0 & + 9.8,- 2.3 & + 0.1,- 0.1 & + 2.1,- 2.6 & + 2.0,- 3.0 & +17.3,-13.8 \\
\nuc{15}{O} &  1.23 & +17.0,-16.0 & +12.9,- 2.3 & + 0.2,- 0.2 & + 2.0,- 2.4 & + 1.9,- 3.0 & +21.5,-16.6 & &  0.77 & +16.0,-15.0 & +12.9,- 2.3 & + 0.2,- 0.2 & + 2.0,- 2.4 & + 1.9,- 3.0 & +20.7,-15.6 \\
\nuc{17}{F} &  0.03 & +19.0,-17.0 & +12.9,- 2.3 & + 0.2,- 0.2 & + 2.0,- 2.4 & + 1.9,- 3.0 & +23.1,-17.6 & &  0.02 & +16.0,-15.0 & +12.9,- 2.3 & + 0.2,- 0.2 & + 2.0,- 2.4 & + 1.9,- 3.0 & +20.8,-15.6 \\
\nuc{8}{B}  &  4.64 & +11.0,-11.0 & +31.8,-14.4 & + 0.5,- 0.4 & + 5.4,- 3.9 & + 1.8,- 2.8 & +34.1,-18.7 & &  3.68 & +11.0,-11.0 & +31.8,-14.4 & + 0.5,- 0.4 & + 5.4,- 3.9 & + 1.8,- 2.8 & +34.1,-18.7 \\
$hep$       &  0.02 & +15.0,-15.0 & +32.7,-15.4 & + 0.3,- 0.3 & + 6.2,- 4.5 & + 1.9,- 2.9 & +36.5,-22.2 & &  0.02 & +15.0,-15.0 & +32.7,-15.4 & + 0.3,- 0.3 & + 6.2,- 4.5 & + 1.9,- 2.9 & +36.5,-22.2 \\
\hline
      Total & 66.31 & + 1.9,- 1.9 & + 3.3,- 1.8 & + 0.1,- 0.1 & + 1.5,- 1.8 & + 1.3,- 2.0 & + 4.3,- 3.8 & & 63.16 & + 1.8,- 1.8 & + 3.1,- 1.8 & + 0.1,- 0.0 & + 1.5,- 1.9 & + 1.4,- 2.1 & + 4.1,- 3.8 \\
\hline
\hline
\end{tabular*}
} % end scriptsize
\end{table*}

In Table~\ref{sigma_table} we give values of $\langle P^{ee}_i\rangle$
and $\langle \sigma_i^{\earth} \rangle$ for each neutrino component.
These were calculated assuming three-neutrino mixing to active
neutrinos with parameters from Ref.~\cite{Schwetz08}: $\Delta
\text{m}^2_{12}=(7.65^{+0.23}_{-0.20}) \times 10^{-5} \text{~eV}^2,
\theta_{12}=33.46^{+1.36}_{-1.00} \text{~degrees, and }
\theta_{13}=5.7^{+3.5}_{-5.7} \text{~degrees}$.  The approximate
formulae given in Ref.~\cite{Barger} were used for the survival probability
$P^{ee}_i(E)$.  As we show in Sec.~\ref{variation} there is no
appreciable difference between the day and night capture rates in Ga
and thus regeneration in the Earth was neglected.  The cross sections
$\sigma(E)$ were taken from Appendix~\ref{sigma_app} for Ga and
Ref.~\cite{clcross} for Cl.  The neutrino spectra
$\phi^{\astrosun}_i(E)$ are from Refs.~\cite{gacross} ($pp$, \nuc{13}{N},
\nuc{15}{O}, \nuc{17}{F}), \cite{clcross} (\nuc{8}{B}), and
\cite{website} ($hep$).

Now that all the terms have been calculated, we can use the fluxes in
Table~\ref{solar_flux}, combined with Eqs.~(\ref{rate_equation}) and
(\ref{sigma_earth}), to predict the capture rate in Ga from each of
the solar neutrino components.  The individual rates and the total
rate are given in Table~\ref{rates} for two recent solar models from
Table~\ref{solar_flux}.  For both models there is good agreement
between the calculated total rate and the observed capture rate of
$66.1 \pm 3.1$~SNU.  The major contribution to the uncertainty in the
predicted total rate is from the neutrino capture cross section, with
smaller contributions from the solar model flux, $\theta_{12}$, and
$\theta_{13}$.

In this analysis we have used the cross sections in
Appendix~\ref{sigma_app}, in which the contribution of the two
lowest-lying excited states in \nuc{71}{Ge} has been set to zero.  If
instead we use the original Bahcall cross sections, then the total
rate increases by 1.2~SNU with the GS98 composition and by 1.1~SNU
with the AGS05 composition.  Whatever cross sections are assumed is
thus not a significant factor in the interpretration of the total rate
in the Ga experiment.  The cross sections are, however, of vital
importance in understanding the origin of the unexpectedly low result
in the source experiments.

The attentive reader may be concerned that there is a logical
inconsistency in the argument presented here: the predicted capture
rates we derive for the Ga experiment depend on the neutrino
oscillation parameters, but the measured total rate of the Ga
experiment is itself one of the inputs used to determine the
oscillation parameters.  Although this is true in a strict sense, the
neutrino oscillation parameters derived from a global fit of all
experiments are for all practical purposes independent of the rate in
the Ga experiment.  Rather, the parameter $\theta_{12}$ is principally
determined by the SNO experiment, $\theta_{13}$ by the CHOOZ
experiment, and $\Delta \text{m}^2_{12}$ by the KamLAND experiment.
Although it was not true in the past, the result of the Ga experiment
is at present only a very minor input to the determination of these
parameters.

Incidentally, if we carry out the same analysis for the \nuc{37}{Cl}
solar neutrino experiment, the total calculated rate is
$3.09(1^{+0.094}_{-0.091})$~SNU using fluxes based on GS98 and
$2.53(1^{+0.091}_{-0.089})$~SNU using fluxes based on AGS05.  These
should be compared with the experimental rate of $2.56 \pm 0.16
\text{~(stat)} \pm 0.16 \text{~(syst)}$~SNU \cite{clapj}.

\section{\texorpdfstring{{The \lowercase{\textit{pp}} neutrino flux %
from the Sun}}{The pp neutrino flux from the Sun}}
\label{pp_flux}
% Programs and calculations are in \solarnu\btc\ppflux.

In this section we will use the Ga measurement given in
Eq.~(\ref{Ga_result}) and the results of other solar neutrino
experiments to determine the $pp$ flux from the Sun.  The conventional
way to make this calculation is by a combined fit to all experiments,
as for example is presented in Ref.~\cite{GGM} and \cite{borexino08}.  Here
we give an alternate approach that successively decomposes the total
measured rate into the components from each neutrino source.  The
final result is identical to what one obtains in a combined fit and
has the advantage that the argument is simple and transparent.

% Strictly speaking, this cannot really be done with the data now
% available as the capture rate in the Ga experiment has contributions
% from all the components of the solar neutrino spectrum, and thus it is
% only possible to separate out the portion from $pp$ when other
% experiments have measured the electron-neutrino part of all the other
% flux components.  Nevertheless, the $pp$ flux can be obtained from the
% present experiments if we make some reasonable approximations.

The rate in Eq.~(\ref{Ga_result}) is the sum of the rates from all the
components of the solar neutrino flux, which we denote by
\begin{equation}
\text{[$pp$+$^7$Be+CNO+$pep+^8$B$|$Ga] = $66.1 (1 \pm 0.047)$ SNU.}
\label{Ga_result2}
\end{equation}
We ignore the tiny $hep$ contribution and combine the \nuc{13}{N},
\nuc{15}{O}, and \nuc{17}{F} components into a single value, called
here ``CNO''.

In an experiment of great technical difficulty, the \nuc{7}{Be} flux
has been directly measured by Borexino and they report the result as
$\phi^{\astrosun}_{\mnuc{7}{Be}} = 5.18(1 \pm 0.098) \times 10^9$
neutrinos/(cm$^2$~s) \cite{borexino08}.  Using
Eqs.~(\ref{rate_equation}) and (\ref{sigma_earth}), we multiply this
flux by the electron neutrino survival factor for \nuc{7}{Be} and by
the cross section of \nuc{7}{Be} on Ga (the values of these factors
and their uncertainties are given in Table~\ref{sigma_table}) and
obtain the rate of \nuc{7}{Be} in Ga of
\begin{equation}
\text{[$^7$Be$|$Ga] = $19.1 (1 ^{+0.12}_{-0.11})$ SNU.}
\end{equation}

The \nuc{8}{B} flux at the Earth has been directly measured by SNO to
be $\phi^{\earth}_{\mnuc{8}{B}} = (1.67 \pm 0.05) \times 10^6$
electron neutrinos/(cm$^2$~s) \cite{SNO08}.  In a similar way to
\nuc{7}{Be}, we multiply this flux by the spectrum-integrated cross
section for \nuc{8}{B} neutrinos on Ga and obtain the \nuc{8}{B}
contribution to the Ga experiment of
\begin{equation}
\text{[$^8$B$|$Ga] = $3.6 (1^{+0.32}_{-0.16})$ SNU.}
\end{equation}

%later
Subtracting these measured rates of \nuc{7}{Be} and \nuc{8}{B}
from the total Ga rate in Eq.~(\ref{Ga_result2}) gives
\begin{equation}
\text{[$pp$+CNO+$pep|$Ga] = $43.3(1^{+0.087}_{-0.094})$ SNU.}
\label{rate_low_plus_medium}
\end{equation}

We can obtain an approximate value for the contribution of CNO and
$pep$ to the Ga experiment from the measured capture rate in the Cl
experiment [$^7$Be+CNO+$pep+^8$B$|$Cl] = $2.56 (1 \pm 0.088)$~SNU
\cite{clapj}.  As in the case of Ga, we use the \nuc{7}{Be} flux
measured by Borexino, the \nuc{8}{B} flux measured by SNO, and the
cross sections in Table~\ref{sigma_table} to determine [$^7$Be$|$Cl] =
$0.67 (1^{+0.105}_{-0.108}$)~SNU and [$^8$B$|$Cl] = $1.73
(1^{+0.068}_{-0.067}$)~SNU.  We subtract these values from the total
Cl rate and are left with [CNO+$pep|$Cl] = $0.19 (1
^{+1.36}_{-1.00}$)~SNU.

If we attribute this entire rate to the neutrinos from $pep$ then,
using the cross sections for $pep$ on Cl and Ga, we calculate a rate
of $[pep|$Ga$]_{\text{test}} = 2.35 (1^{+1.37}_{-1.00})$~SNU.  On the
other hand, if we attribute this entire rate to CNO, we obtain in the
same manner a rate of [CNO$|$Ga$]_{\text{test}} = 3.11
(1^{+1.37}_{-1.00})$~SNU.  The upper extreme of these two test rates
is $3.11 \times (1 + 1.37) = 7.37$~SNU.  As a reasonable estimate we
can thus set the sum of CNO and $pep$ rates at half this upper limit
with an uncertainty of 100\%:
\begin{equation}
\text{[CNO$+pep|$Ga] = $3.68 (1^{+1.00}_{-1.00})$ SNU.}
\end{equation}

We subtract this estimate for the CNO plus $pep$ rate from the rate in
Eq.~(\ref{rate_low_plus_medium}) and obtain the result for the
measured $pp$ rate in the Ga experiment
\begin{equation}
\text{$[pp|$Ga] = $39.7 (1^{+0.13}_{-0,14})$ SNU.}
\end{equation}
Dividing this capture rate by the cross section for capture of $pp$
neutrinos from Table~\ref{sigma_table} gives the measured electron
neutrino $pp$ flux at Earth of
\begin{equation}
\text{$\phi^{\earth}_{pp} = 3.38 (1^{+0.14}_{-0.14}) \times 10^{10}$/(cm$^2$~s).}
\label{pp_flux_earth}
\end{equation}
If we use Eq.~(\ref{flux_ratio}) and the value of $\langle
P^{ee}_i\rangle = 0.561 (1 ^{+0.030} _{-0.042})$ from
Table~\ref{sigma_table} then the $pp$ flux produced in the Sun is
\begin{equation}
\text{$\phi^{\astrosun}_{pp}=6.0 (1 \pm 0.14) \times 10^{10}$/(cm$^2$~s).}
\label{pp_flux_sun}
\end{equation}
Our present result for the $pp$ flux is in good agreement with the
previous estimates that we have made during the past six years
\cite{jetp,taup03,neut08}, with the major change being a gradual
reduction of the uncertainty.  In the future, as Borexino continues to
collect data, and as direct measurements are made of the CNO and $pep$
fluxes, the uncertainty in this flux should be further reduced and
eventually may be dominated by the uncertainty in the Ga rate itself.  By
that time, however, there will hopefully be direct experiments that
measure the $pp$ flux in real time.

For comparison, we see from Table~\ref{solar_flux} that the predicted
$pp$ flux from the two recent solar models with different composition
is $\phi^{\astrosun}_{pp}=5.97 \pm 0.04$ and $6.04 \pm 0.03$, both in
units of $10^{10}~\nu_e$/(cm$^2$~s).  There is good numerical
agreement between these flux values and the result in
Eq.~(\ref{pp_flux_sun}), but, as made clear by Bahcall and
Pe\~na-Garay \cite{road_map}, there is a large difference in
interpretation: the result in Eq.~(\ref{pp_flux_sun}) was derived from
contemporary solar neutrino experiments and is the $pp$ neutrino flux
\textit{at the present time}.  In contrast, energy generation in the
solar model is highly constrained by the measured solar luminosity and
thus, when the luminosity constraint is imposed, as is the case for
the models given in Table~\ref{solar_flux}, the calculated $pp$ flux
is what the Sun was producing some $40~000$ years ago.  The agreement
between the present $pp$ flux, as measured by the Ga experiment, and
the past flux, as inferred from the solar model with the luminosity
constraint, implies that the $pp$ flux from the Sun has not altered
(within our 14\% uncertainty) during the last $40~000$ years.

% for approximation to Pee look at Gonzalez-Garcia + Maltoni p. 121

\section{Consideration of time variation}
\label{variation}

In a plot of the SAGE results as a function of time there is a slight
visual hint of a long-term decrease, as illustrated in
Fig.~\ref{yearly}.  The average rate prior to 1996 is somewhat
higher than after 1996.  A plot of the Gallex-GNO data shows a similar
behavior \cite{gno_final}.  When examined quantitatively, however, the
evidence for a long-term decrease in the capture rate is unconvincing.
A $\chi^2$ test applied to these yearly SAGE data points assuming the
rate is constant at 65.4~SNU gives $\chi^2$/DOF~=~12.0/17, which has a
probability of 80\%.  The fit to a constant rate is thus quite good.

\begin{figure}
\centering

\ifpdf
 \includegraphics[width=\hsize,viewport=21 13 437 285]{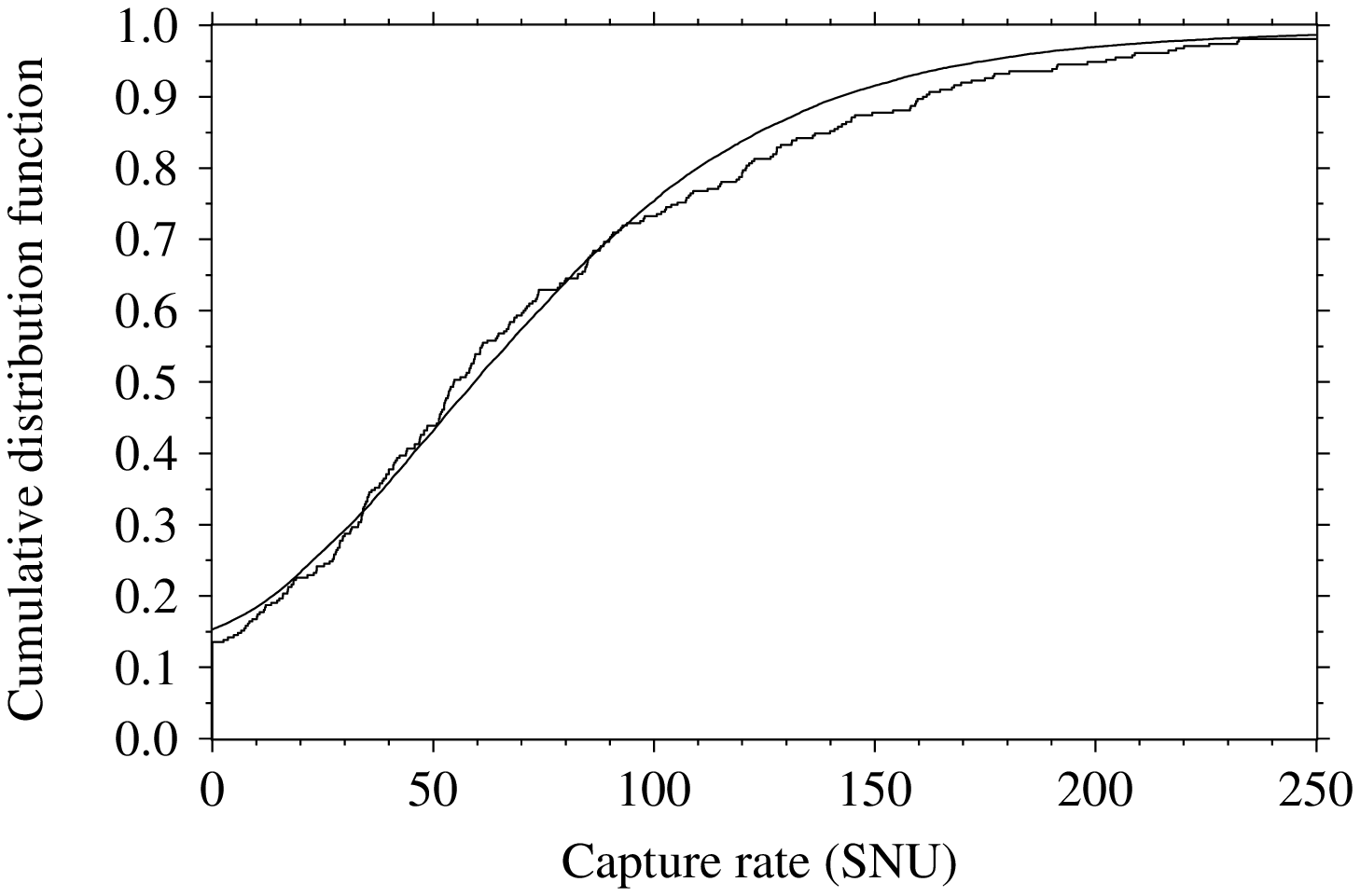}
\else
 \includegraphics[width=\hsize,bb=21 13 437 285]{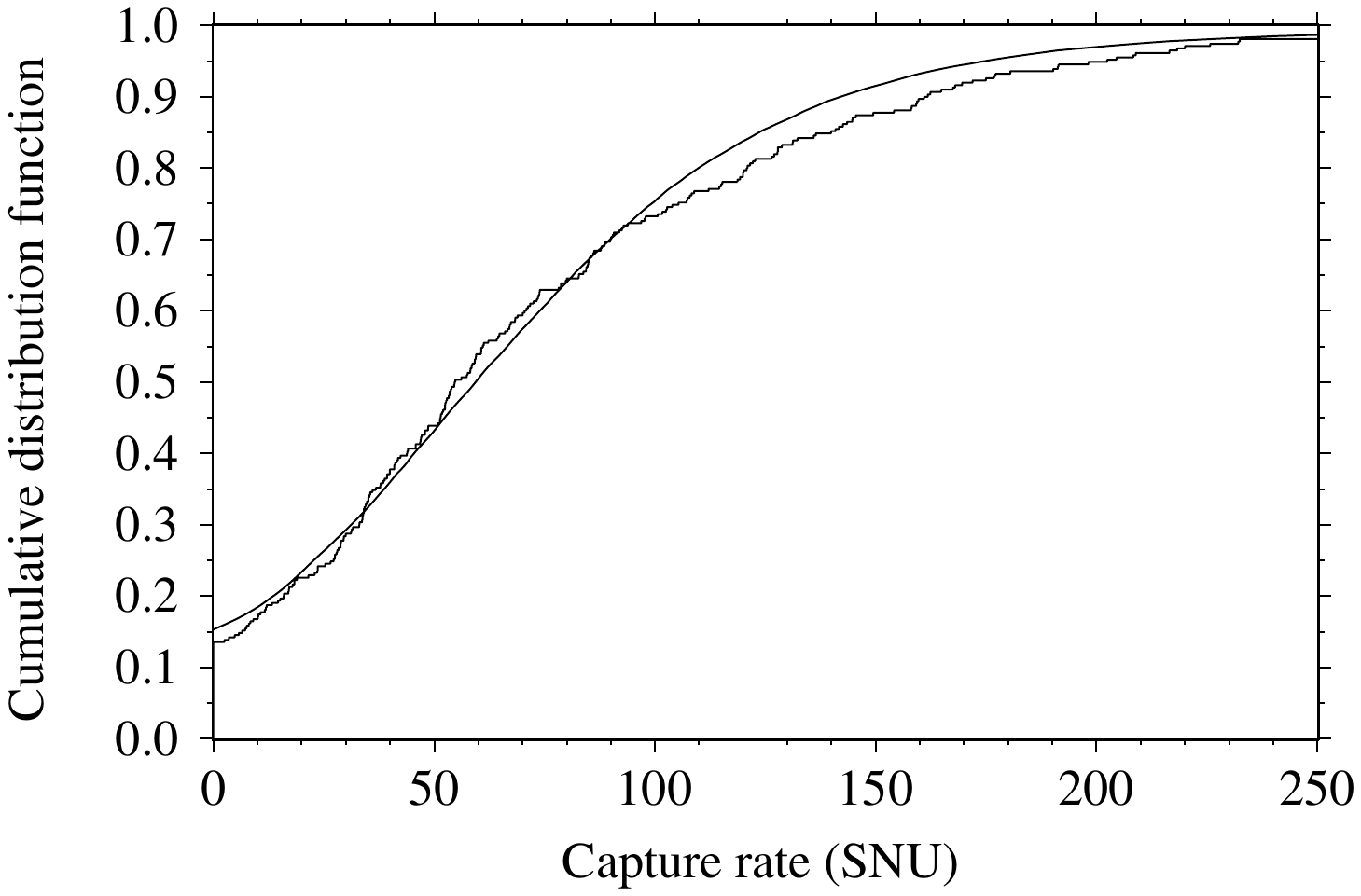}
\fi

\caption{Measured capture rate for all 310 SAGE data sets (jagged
curve) and the expected distribution derived by 100 Monte Carlo
simulations of each set (smooth curve).}

\label{prod}
\end{figure}

In previous articles we have demonstrated the agreement between the
assumption of a constant production rate and the SAGE measurements by
use of the cumulative distribution of the capture rate $C(p)$,
defined as the fraction of data sets whose capture rate is less
than $p$.  Figure \ref{prod} shows this distribution for the data and
the expected distribution derived from 100 simulations of all
168~runs, where it is assumed in the simulations that the production
rate is constant and has a value of 65.4~SNU.  For each run the rates
from the separate $L$ and $K$ peaks are used in this figure, not the
rate from the $L+K$ combination.  To ensure that the simulations
parallel the real data as closely as possible, all parameters of the
simulation, such as background rates, efficiencies, exposure times,
and counting times, were chosen to be the same as for the real data.
Only the number of counts in each run and the times when these counts
occurred were allowed to vary.

The data spectrum and the simulated spectrum are very similar to each
other, indicating that the distribution of capture rates is what one
would expect if the rate is constant.  A quantitative comparison can
be made by calculating the $Nw^2$ test statistic for the data
distribution and comparing it to the distribution from simulations
using the method described in Ref.~\cite{Cleveland98}.  The fraction of
simulated spectra whose $Nw^2$ was larger than for the data
distribution is 43\%, which shows that the assumption of a constant
capture rate is in good agreement with our measurements.

\begin{figure}
\centering

\ifpdf
 \includegraphics[width=\hsize,viewport=3 19 583 439]{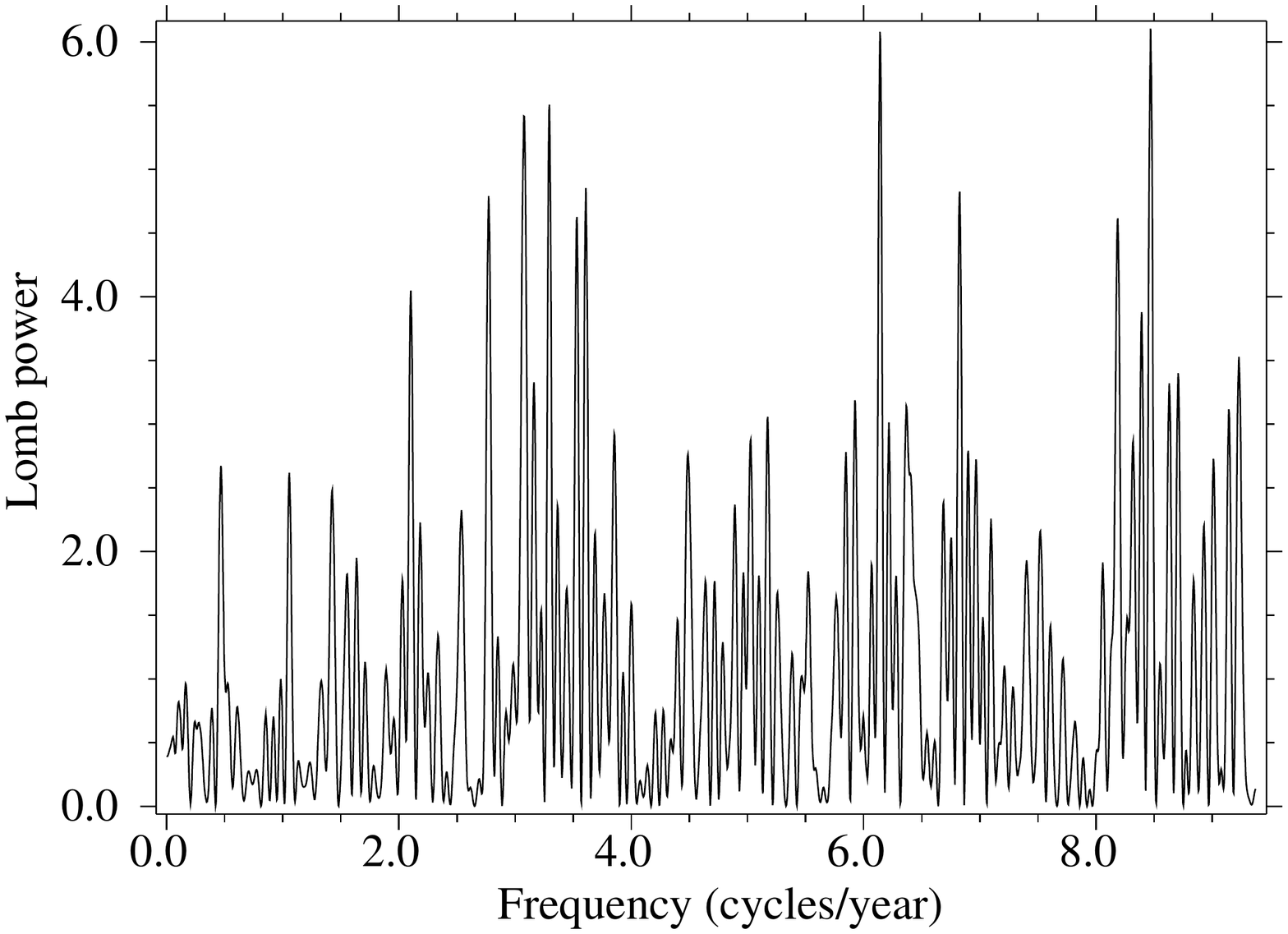}
\else
 \includegraphics[width=\hsize,bb=3 19 583 439]{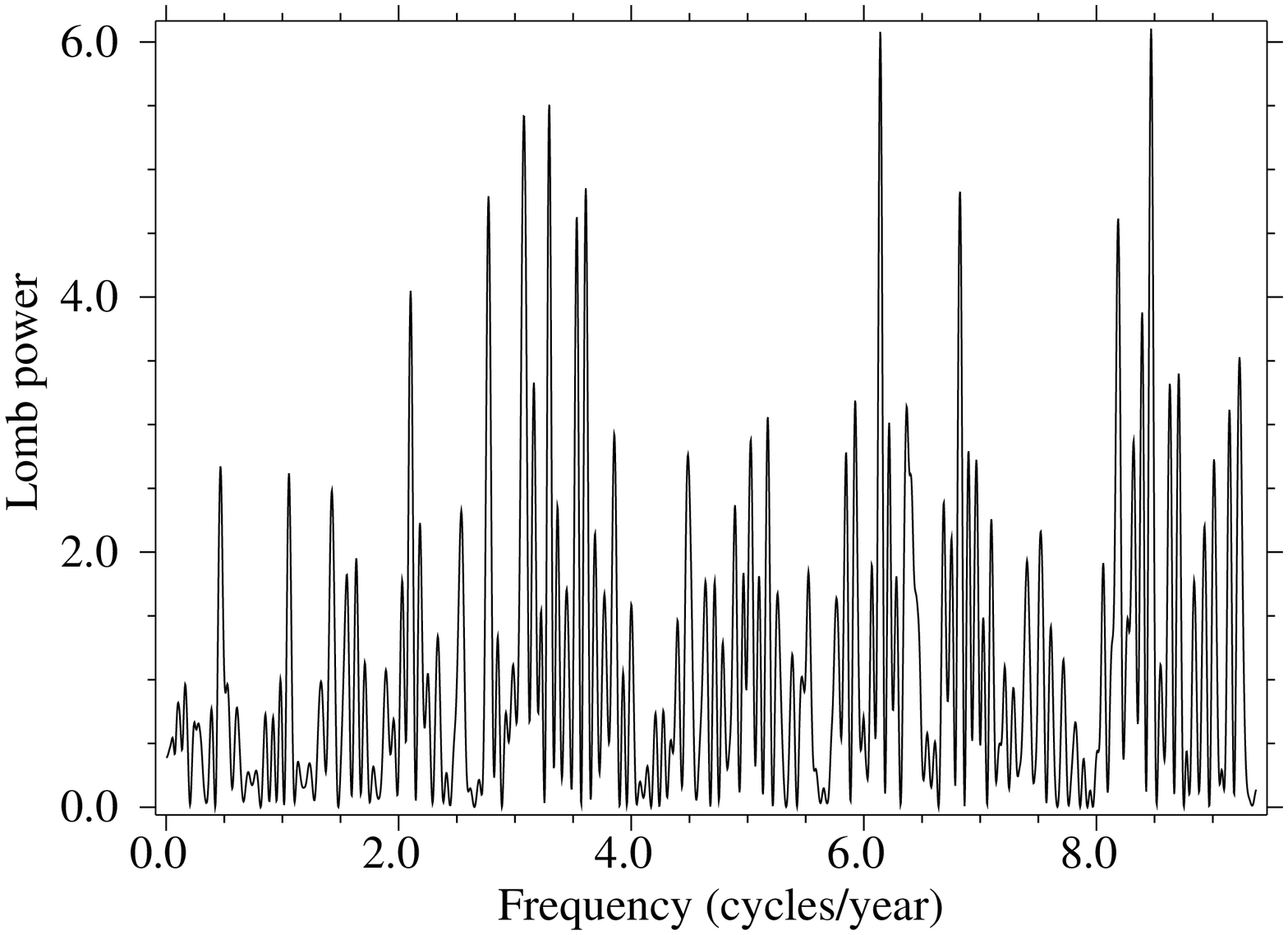}
\fi

\caption{Lomb power spectrum from all 168~SAGE data runs.  The mean
time of exposure was used as the time of measurement.}

\label{Lomb_power}
\end{figure}

A standard method to look for periodic signals in unevenly sampled
data, such as we have in SAGE, was devised by Lomb and Scargle.
Application of this method, using the implementation of Press
\textit{et al.\/}~\cite{numrep}, to all runs from the SAGE experiment
yields the power spectrum shown in Fig.~\ref{Lomb_power}.  The
frequency range considered is from nearly zero up to slightly less
than twice the Nyquist frequency.  The maximum Lomb power is 6.10 and
it occurs at a frequency of 8.47~cycles/year.

\begin{figure}
\centering

\ifpdf
 \includegraphics[width=\hsize,viewport=24 25 584 439]{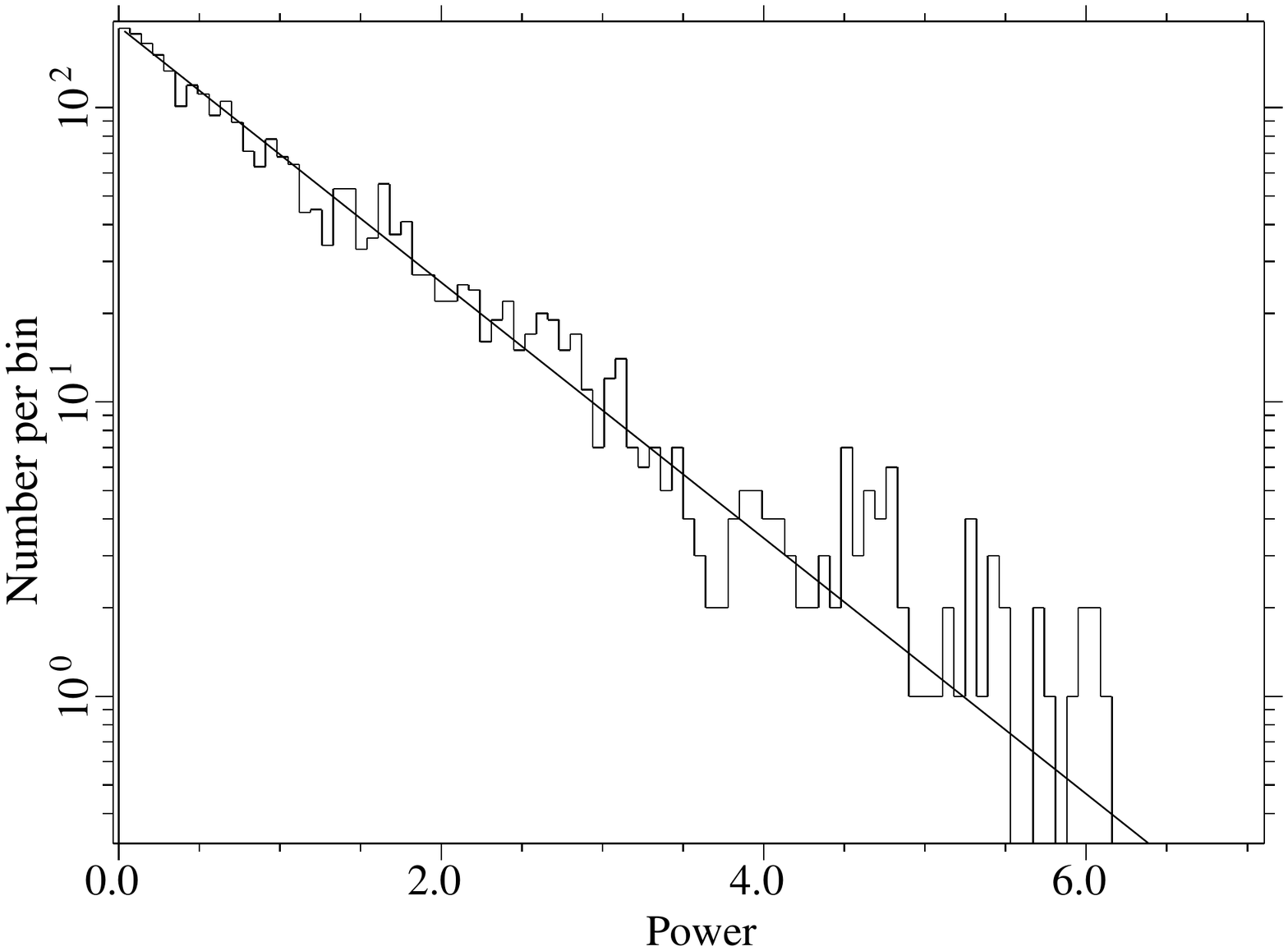}
\else
 \includegraphics[width=\hsize,bb=24 25 584 439]{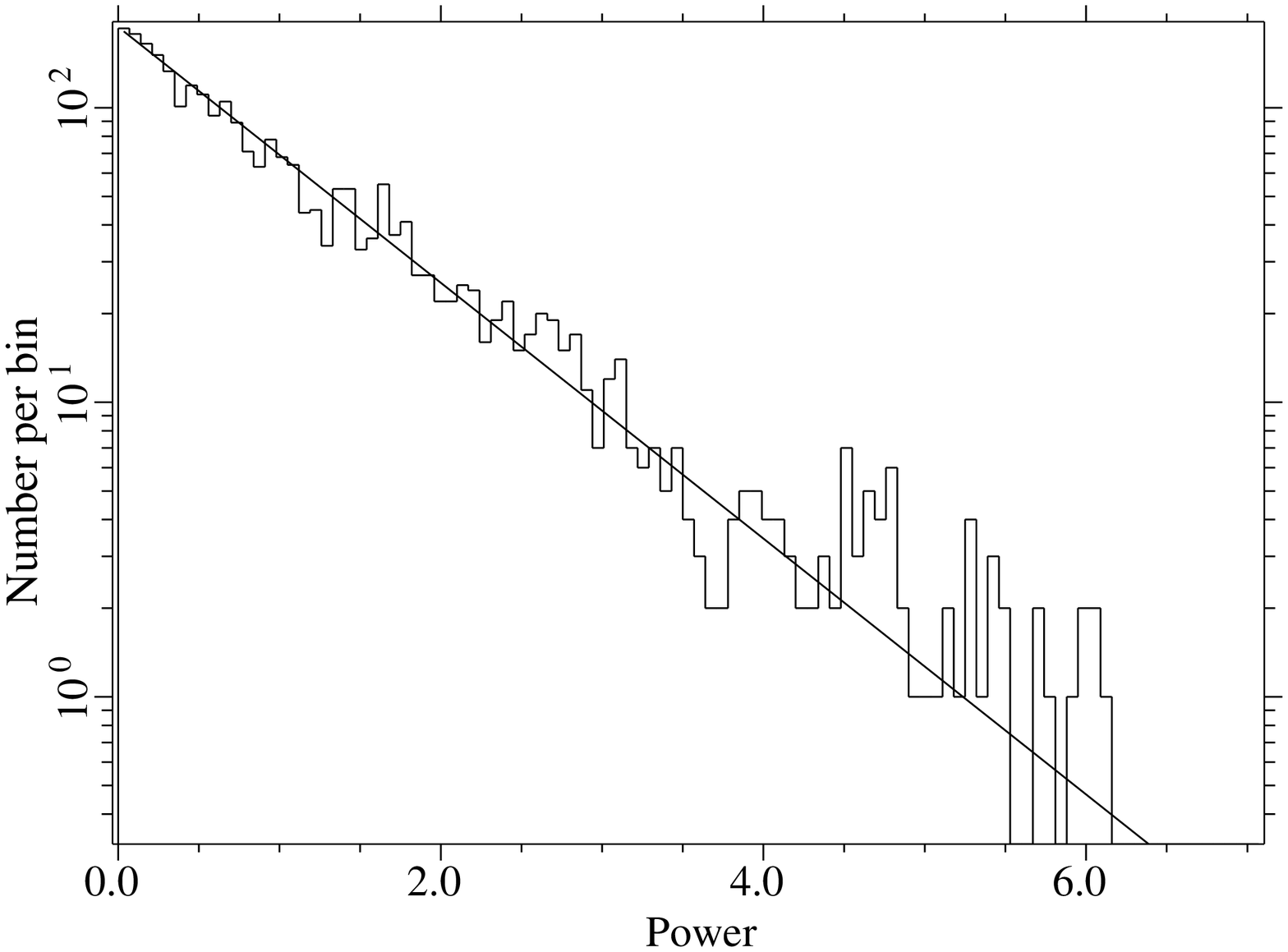}
\fi

\caption{Histogram of powers in spectrum of
Fig.~\protect{\ref{Lomb_power}}.  The bin size is 0.07~power units.
The solid line is the expected distribution if there is no time
variation, i.e., the $\text{number of frequencies} \times
\exp(-\text{power})$, integrated over the limits of each bin.}

\label{power_distribution}
\end{figure}

A simple way to assess the significance of a peak in such a spectrum
is to make a histogram of the number of frequencies as a function of
power.  In the absence of any time variation this distribution is an
exponential; if there were any peak present with significant power it
would appear at the upper end of the distribution and be clearly
separated from the exponential trend.  This distribution for the
spectrum of Fig.~\ref{Lomb_power} is shown in
Fig.~\ref{power_distribution}.  As this distribution visually shows,
there is no evidence for exceptionally high power in the data spectrum
at any frequency.

\begin{figure}
\centering

\ifpdf
 \includegraphics[width=\hsize,viewport=3 18 582 439]{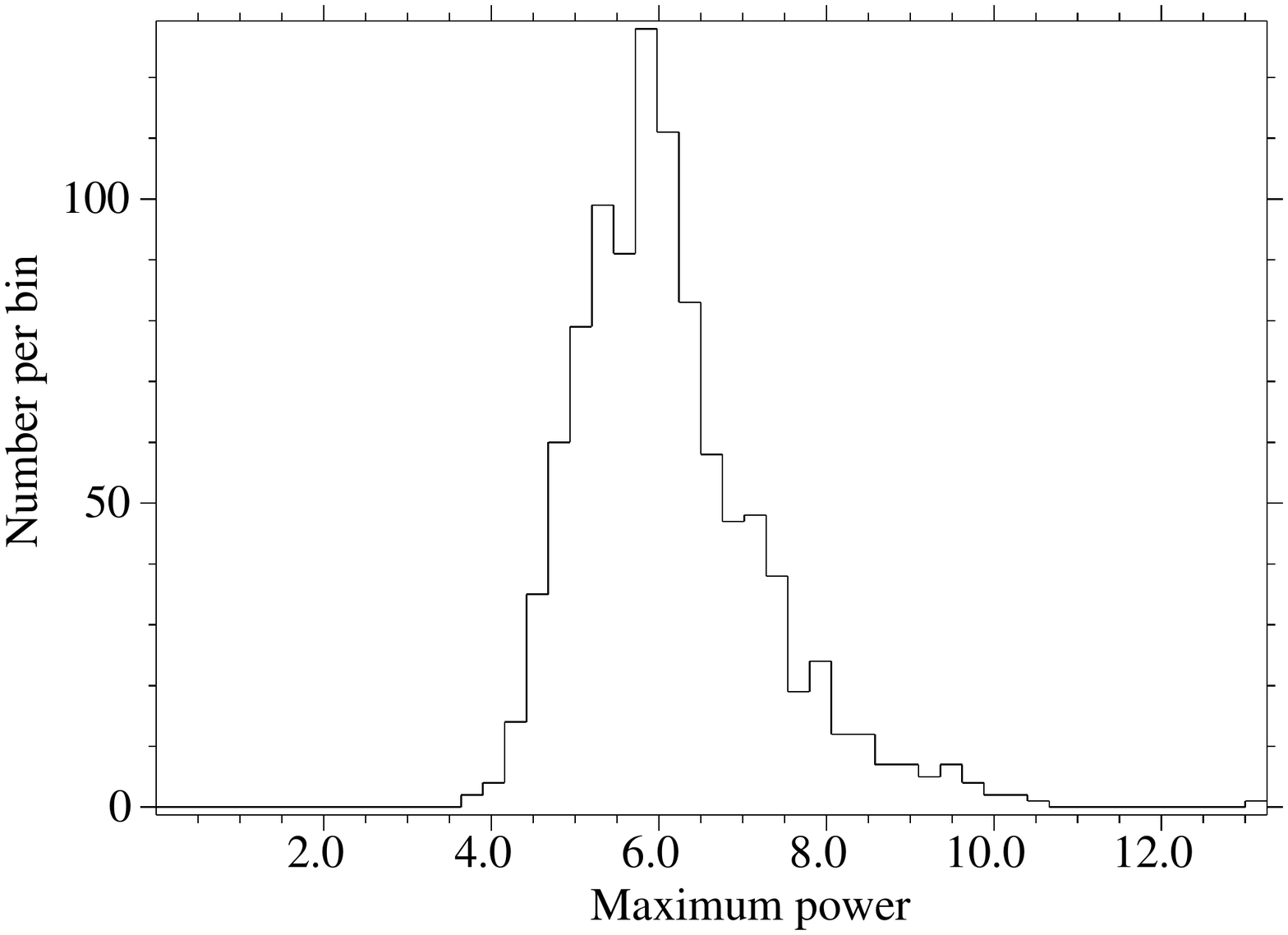}
\else
 \includegraphics[width=\hsize,bb=3 18 582 439]{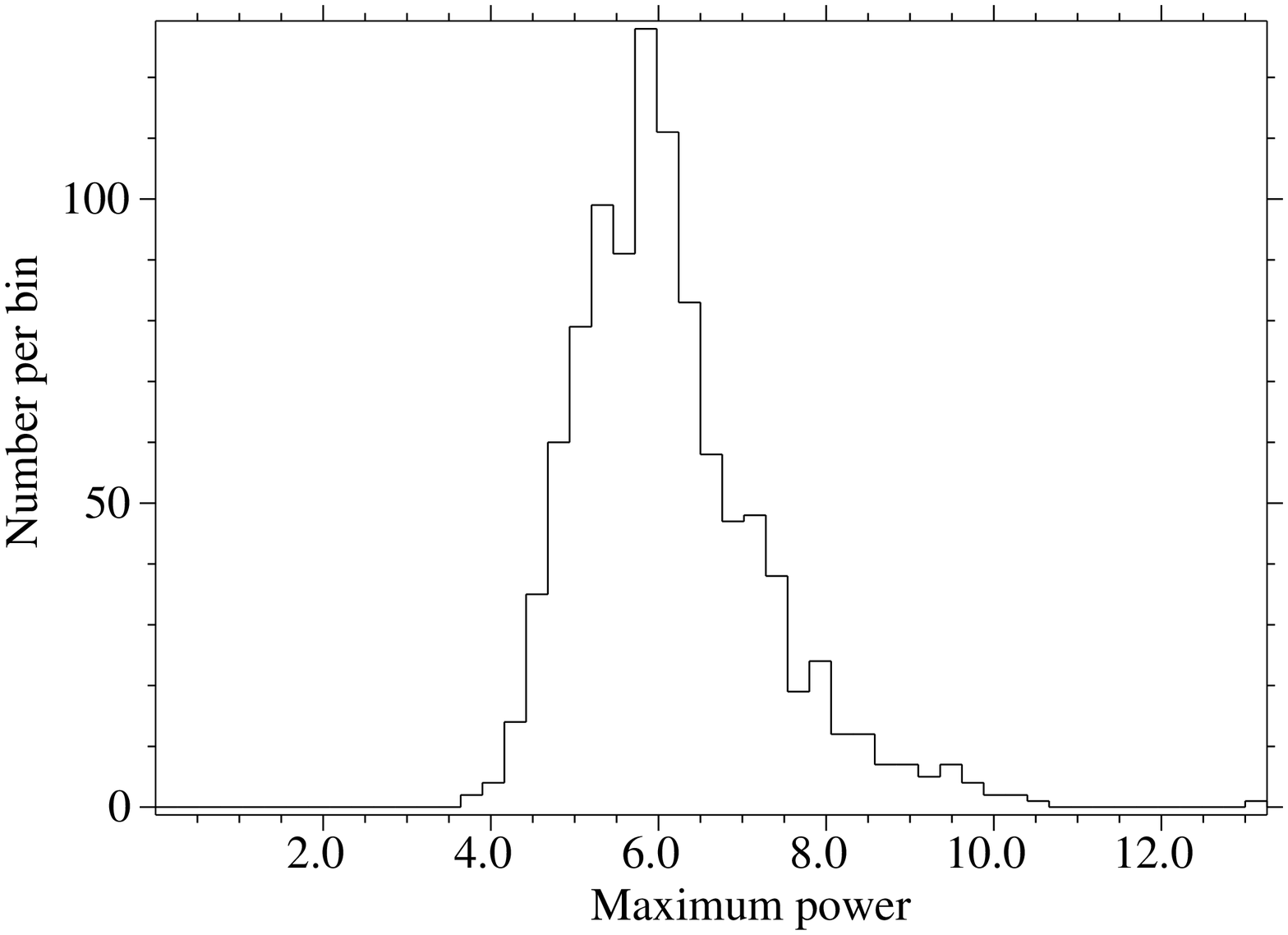}
\fi

\caption{Histogram of maximum power found in Lomb power spectrum
analysis of 1000 random shuffles of the 168~SAGE runs.  The bin size
is 0.13~power units.}

\label{shuffle_power}
\end{figure}

A quantitative way to prove that no frequency has exceptionally high
power is with a shuffle test.  In this test the SNU results are
randomly re-assigned to the different runs, the power spectrum is
recalculated, and the maximum power is found.  The maximum power in
the spectra from 1000 such shuffles is plotted in
Fig.~\ref{shuffle_power}.  The observed maximum power for the SAGE
data of 6.10 occurs very near the center of this distribution.
Of the shuffles, 44\% have a greater power than for the data and 56\% a
lesser power, thus showing that the observed power distribution is
consistent with the assumption of a constant rate.

The same result is obtained if power spectra are produced over a wider
frequency range.  For ranges up to 19~per year and 38~per year the
maximum power remains 6.10 at 8.47~cycles per year and is not
statistically significant.

The three tests in this section do not find any evidence for periodic
variation in the SAGE data.  The frequency range to which these tests
are sensitive is about one cycle per month to about one cycle per ten
years.  For a frequency to be detected within this range the amplitude
of the periodic oscillation must, as shown by Pandola \cite{Pandola},
exceed the statistical uncertainty of a single run, i.e., must be
$\gtrsim$ 30~SNU.

We end this section by noting that the winter minus summer difference
in SAGE capture rate is $R_{\text{W}} - R_{\text{S}} =
5.8^{+6.2}_{-6.1}$~SNU where the stated uncertainty is only
statistical.  For this calculation summer was defined as the
$\pm$\textonequarter-year interval centered on 21~June and winter as
the rest of the year.  In our method of data analysis \cite{prc} we
remove the known change in rate caused by the Earth's orbital
eccentricity.  If, rather than using the above solstice-based
definition, we define summer as the $\pm$\textonequarter-year
interval centered on 5~July, the time of the aphelion, then
$R_{\text{W}} - R_{\text{S}} = 4.2^{+6.2}_{-6.1}$~SNU.  With both of
these definitions $R_{\text{W}} - R_{\text{S}}$ is consistent with
zero, indicating that there is no appreciable difference between the
day and night capture rates in Ga, as is expected for the currently
determined values of the neutrino oscillation parameters
\cite{Fogli99}).

\section{Summary and Discussion}

In 18~years of operation the SAGE experiment has carried out
168~measurements of the solar neutrino capture rate.  Analysis of the
times of occurrence of all events in windows centered on the $L$ and
$K$ peaks and with 2~FWHM width ascribes 853.9 of these events to the
decay of \nuc{71}{Ge}.  To compare these values with other
radiochemical solar neutrino experiments, the Cl experiment collected
solar data for a longer time (108 runs with rise-time counting during
24~years), but the number of detected events within an energy window
of 2~FWHM that were ascribed to \nuc{37}{Ar} was 875 \cite{clapj},
quite comparable to what we now have in SAGE.  So far as we are aware,
neither Gallex nor GNO have reported their total number of detected
\nuc{71}{Ge} events, but because the mass of their Ga target was less
(30 tonnes), and their number of runs was less (123 during 12~years),
this number must be considerably less than we now have in SAGE.

The measured best-fit capture rate in the SAGE experiment is $65.4
^{+3.1}_{-3.0}$ (stat) $^{+2.6}_{-2.8}$ (syst)~SNU.  Combining this
with the results of Gallex and GNO gives the rate as $66.1 \pm
3.1$~SNU, where statistical and systematic uncertainties have been
combined in quadrature.  The solar-model prediction for the Ga
experiment is 66.3~SNU for a model with high metallicity and 63.2~SNU
for a model with low metallicity, where the uncertainty of both
estimates is $\sim$4\%.  There is thus excellent agreement between
theory and experiment for the Ga experiment.  Further, both the
experimental measurement and the theoretical prediction are known to
about the same accuracy.

By use of the results of other solar neutrino experiments and neutrino
oscillation theory we have derived the contemporary value of the $pp$
flux from the Sun to be $(3.40^{+0.46}_{-0.47}) \times
10^{10}$/(cm$^2$~s) at the Earth and $(6.0 \pm 0.8) \times
10^{10}$/(cm$^2$~s) at the Sun.  The latter is in good agreement with
standard solar model predictions of $5.97 \pm 0.04$ (high metallicity)
and $6.04 \pm 0.03$ (low metallicity), both in units of
$10^{10}~\nu_e$/(cm$^2$~s).  Gallium experiments have thus proven that
the overwhelming fraction of solar neutrinos that reach the Earth are
the low-energy neutrinos from the $pp$~reaction.

We have assumed in these calculations that the cross section for
capture to the two lowest-lying excited states in \nuc{71}{Ge} is
zero, as is implied by the four neutrino source experiments with
gallium.  This assumption is in contradiction to the standard
interpretation of the two experiments that have attempted to measure
the Gamow-Teller strength of these low-lying excited states in
\nuc{71}{Ge}.  These experiments were made by $(p,n)$ scattering
\cite{gacross,pnscattering} and (\nuc{3}{He},$t$) scattering
\cite{Ejiri1,Ejiri2,Hetscattering}.  If all the events observed in
these experiments at low excitation energy are attributed to
Gamow-Teller strength the results of these experiments are in
reasonably good agreement.  It is, however, not evident that these
experiments solely measure Gamow-Teller strength--as emphasized by
Haxton \cite{Haxtoncross}, for very weak transitions, such as is the
case in these experiments, there may be an appreciable (perhaps
dominant) contribution to the cross section from the spin-tensor
interaction.  New experimental data is needed to settle this question
and to definitively determine the magnitude of the matrix elements for
neutrino capture to these two low-lying excited states.  We strongly
encourage any new experiments that might shed light on this question.
As part of our future experimental program we intend to pursue a new
measurement that will use a very intense neutrino source in an
optimized detector geometry.

The SAGE experiment continues to collect data on the solar neutrino
capture rate with a gallium target.  Up to now it is only the Ga
experiment that has measured the low-energy $pp$ solar neutrinos.  As
we continue to monitor the solar neutrino flux we will increase our
statistical accuracy and further reduce our systematic uncertainties.

\begin{acknowledgments} 

SAGE is grateful to M. Baldo-Ceolin, W. Haxton, V. A. Kuzmin,
V. A. Matveev, S. P. Mikheev, R. G. H. Robertson,
V. A. Rubakov, A. Yu. Smirnov, A. Suzuki, A. N. Tavkhelidze,
and our colleagues from the GALLEX and GNO collaborations for their
continued interest and for fruitful and stimulating discussions.  We
especially thank W. Hampel for vital comments on many aspects of our
investigations.

SAGE acknowledges the support of the Russian Academy of Sciences, the
Ministry of Education and Science of the Russian Federation, the
Division of Nuclear Physics of the US~Department of Energy, and
the US~National Science Foundation.

This work was partially funded by the Russian Foundation for Basic
Research under grants 99-02-16110, 02-02-16776,
05-02-17199 and 08-02-00146, by the Program of the President
of the Russian Federation under grants 00-1596632,
NS-1782.2003, NS-5573.2006.2 and NS-959.2008.2, by the
Program of Basic Research ``Neutrino Physics'' of the Presidium of the
Russian Academy of Sciences, by the International Science and
Technology Center under grant 1431, and by the US~Civilian
Research and Development Foundation under grants CGP~RP2-159 and
CGP~RP2-2360-MO-02.

\end{acknowledgments}

\appendix
\section{Calculation of extraction efficiency from isotopic analysis}
\label{ext_eff_calc}

It is assumed the extracted Ge consists of a combination of
\begin{itemize}
\item Ge from the carrier added for the current extraction,
\item residual Ge that remained from the carrier added for the two
      preceding extractions, and
\item additional Ge with natural isotopic composition, such as may be
      dissolved from the surfaces of the extraction system or the
      vessels that contain the Ga.
\end{itemize}

According to this model the predicted mass of isotope $i$ obtained in
extraction $n$, called $M_n^p(i)$, is thus
\begin{equation}
M_n^p(i)\! =\! \varepsilon_n \{C_n(i) + (1-\varepsilon_{n-1}) [C_{n-1}(i) + (1-\varepsilon_{n-2}) C_{n-2}(i)]\} + E_n I(i),
\end{equation}
\noindent where $\varepsilon_m$ is the efficiency of Ge removal in
extraction $m$, $C_m(i)$ is the mass of isotope $i$ of carrier added
to extraction $m$ (where $m$ can take on the values $n$, $n-1$, or
$n-2$), and $E_n$ is the mass of additional Ge with natural isotopic
composition $I(i)$ that is removed in extraction $n$.  It is assumed
that $I(i), C_n(i), C_{n-1}(i),$ and $C_{n-2}(i)$ are known and the
variables $\varepsilon_n, \varepsilon_{n-1}, \varepsilon_{n-2},$ and
$E_n$ are to be determined.  For $N$ extractions there are thus $2N$
variables (the extraction efficiency and mass of extra natural Ge for
each extraction) and $5N$ equations that relate these variables, one
for each of the naturally-occurring Ge isotopes, \nuc{70}{Ge},
\nuc{72}{Ge}, \nuc{73}{Ge}, \nuc{74}{Ge}, and \nuc{76}{Ge}.  Because
there are more relationships than unknowns, the problem is solved by
finding the set of variables that minimizes the function
\begin{equation}
\chi^2 = \sum_{n=1}^N \sum_{i=1}^5 \biggl( \frac{M_n^p(i) - M_n^e(i)}{\sigma_n(i)} \biggr)^2,
\end{equation}
\noindent where $M_n^e(i)$ is the measured mass of isotope $i$ in
extraction $n$ and $\sigma_n(i)$ is the total uncertainty in the
knowledge of the predicted and measured masses in extraction $n$.

\section{Results for each SAGE extraction}
\label{results_table}

The capture rate for each SAGE extraction is given in
Table~\ref{table_of_results}.  The statistic $Nw^2$ in this table
measures the goodness of fit between the observed sequence of events
in time and the time distribution predicted by the model used in
analysis, viz., that the events are produced by the sum of two
processes: the decay of a fixed initial number of \nuc{71}{Ge} atoms
and background events at a constant rate.  The probability to obtain a
value of $Nw^2$ larger than in the observed time distribution is given
in the last column.  It is derived by 1000 simulations for each
extraction using the method in Ref.~\cite{Cleveland98} and has an
uncertainty of $\sim$1.5\%.  The time that the exposure interval for
each extraction began and ended can be obtained from the entries in
columns 2 and 3 by use of transformation equations (6.4) in
Ref.~\cite{prc}.

\begin{longtable*}{@{} l @{\extracolsep{\fill}} c d d d d d r @{--} @{\extracolsep{0em}} d @{\extracolsep{\fill}} d d @{}}

\caption{Results of analysis of SAGE extractions.} \rule{0pt}{5ex} \\

\label{table_of_results}

           & Median   & \mco{Exposure} & \mco{Ga}     & \mco{Number of} & \mco{Number}   \\
Exposure   & exposure & \mco{time}     & \mco{mass}   & \mco{candidate} & \mco{fit to}     & \mco{Best fit} & \multicolumn{2}{c}{68\% conf.}   &              & \mco{Probability} \\
 date      & date     & \mco{(days)}   & \mco{(tons)} & \mco{events}    & \mco{$^{71}$Ge}  & \mco{(SNU)}    & \multicolumn{2}{c}{range (SNU)}  & \mco{$Nw^2$} & \mco{(\%)}        \\
\hline
 Jan. 90   & 1990.040 &  42.0 & 28.67 &     8 &    0.0 &    0 &    0 &   65 & 0.532 &    2 \\
 Feb. 90   & 1990.139 &  30.0 & 28.59 &     2 &    1.5 &   74 &   19 &  160 & 0.167 &   25 \\
 Mar. 90   & 1990.218 &  26.0 & 28.51 &    10 &    1.0 &   40 &    0 &  211 & 0.040 &   83 \\
 Apr. 90   & 1990.285 &  19.0 & 28.40 &    11 &    0.0 &    0 &    0 &  157 & 0.119 &   35 \\
 July 90   & 1990.540 &  21.0 & 21.01 &    13 &    0.0 &    0 &    0 &  252 & 0.080 &   50 \\
\\
 June 91   & 1991.463 &  53.0 & 27.43 &    10 &    0.0 &    0 &    0 &  120 & 0.188 &   20 \\
 July 91   & 1991.539 &  23.0 & 27.37 &     1 &    0.6 &   34 &    0 &  116 & 0.163 &   33 \\
 Aug. 91   & 1991.622 &  26.3 & 49.33 &    16 &    9.4 &  395 &  247 &  584 & 0.036 &   85 \\
 Sep. 91   & 1991.707 &  27.0 & 56.55 &    11 &    2.0 &   42 &    9 &  123 & 0.023 &   97 \\
 Nov. 91   & 1991.872 &  26.0 & 56.32 &    31 &    3.1 &   61 &    9 &  162 & 0.173 &   12 \\
 Dec. 91   & 1991.948 &  26.8 & 56.24 &    10 &    8.8 &  159 &  100 &  219 & 0.061 &   79 \\
\\
 Feb. 92-1 & 1992.138 &  24.5 & 43.03 &    14 &    0.0 &    0 &    0 &   43 & 0.108 &   44 \\
 Feb. 92-2 & 1992.138 &  24.5 & 13.04 &     1 &    0.8 &   80 &    0 &  193 & 0.084 &   87 \\
 Mar. 92   & 1992.214 &  20.9 & 55.96 &    24 &   11.7 &  285 &  203 &  414 & 0.077 &   36 \\
 Apr. 92   & 1992.284 &  23.5 & 55.85 &    15 &    1.4 &   34 &   13 &  112 & 0.143 &   20 \\
 May  92   & 1992.383 &  27.5 & 55.72 &     5 &    0.0 &    0 &    0 &   86 & 0.142 &   33 \\
 Sep. 92   & 1992.700 & 116.8 & 55.60 &    11 &    6.5 &   84 &   52 &  125 & 0.120 &   24 \\
 Oct. 92   & 1992.790 &  27.2 & 55.48 &    18 &    3.3 &   31 &    7 &   63 & 0.093 &   37 \\
 Nov. 92   & 1992.871 &  26.7 & 55.38 &    28 &    6.9 &   90 &   45 &  145 & 0.143 &   13 \\
 Dec. 92   & 1992.945 &  24.3 & 55.26 &    27 &   17.6 &  174 &  121 &  229 & 0.063 &   57 \\
\\
 Jan. 93   & 1993.039 &  32.3 & 55.14 &    17 &    9.9 &  122 &   74 &  176 & 0.093 &   33 \\
 Feb. 93   & 1993.115 &  23.0 & 55.03 &     3 &    0.8 &   18 &    0 &   56 & 0.090 &   47 \\
 Apr. 93   & 1993.281 &  26.6 & 48.22 &     7 &    2.3 &   56 &   15 &  106 & 0.038 &   90 \\
 May  93   & 1993.364 &  30.9 & 48.17 &     8 &    0.6 &   28 &    0 &  122 & 0.115 &   41 \\
 June 93   & 1993.454 &  30.4 & 54.66 &    18 &    5.1 &   63 &   22 &  116 & 0.426 &    0 \\
 July 93   & 1993.537 &  27.9 & 40.44 &    28 &    6.7 &  198 &  100 &  312 & 0.041 &   84 \\
 Aug. 93-1 & 1993.631 &  34.0 & 40.36 &     4 &    2.7 &   73 &   28 &  125 & 0.051 &   81 \\
 Aug. 93-2 & 1993.628 &  63.8 & 14.09 &     1 &    1.0 &  120 &    0 &  230 & 0.093 &   75 \\
 Oct. 93-1 & 1993.749 &  13.0 & 14.06 &     0 &    0.0 &    0 &    0 &  158 & $NA$  & $NA$ \\
 Oct. 93-2 & 1993.800 &  34.7 & 14.10 &     4 &    3.1 &  144 &   71 &  246 & 0.052 &   86 \\
 Oct. 93-3 & 1993.812 &  24.6 & 14.02 &     6 &    2.9 &  132 &   64 &  231 & 0.049 &   82 \\
\\
 July 94   & 1994.551 &  31.3 & 50.60 &    20 &    4.5 &   63 &   29 &  108 & 0.018 &  100 \\
 Aug. 94   & 1994.634 &  31.0 & 50.55 &    25 &    3.6 &   42 &   14 &   79 & 0.031 &   95 \\
 Sep. 94-1 & 1994.722 &  33.2 & 37.21 &    30 &    5.9 &  101 &   42 &  174 & 0.100 &   36 \\
 Oct. 94   & 1994.799 &  28.8 & 50.45 &    44 &    0.0 &    0 &    0 &  128 & 0.269 &   12 \\
 Nov. 94   & 1994.886 &  31.0 & 50.40 &    23 &    8.0 &  115 &   68 &  172 & 0.015 &  100 \\
 Dec. 94   & 1994.951 &  21.0 & 13.14 &     9 &    0.0 &    0 &    0 &  236 & 0.184 &   20 \\
\\
 Mar. 95   & 1995.209 &  42.5 & 24.03 &    23 &    3.6 &  145 &   48 &  264 & 0.042 &   84 \\
 July 95   & 1995.538 &  19.9 & 50.06 &    33 &    7.3 &  106 &   53 &  168 & 0.108 &   28 \\
 Aug. 95   & 1995.658 &  46.7 & 50.00 &    21 &    7.5 &  105 &   62 &  158 & 0.081 &   43 \\
 Sep. 95   & 1995.742 &  28.8 & 49.95 &    33 &    1.3 &   29 &    0 &  126 & 0.058 &   75 \\
 Oct. 95   & 1995.807 &  18.7 & 49.83 &    25 &    5.8 &  148 &   62 &  254 & 0.037 &   89 \\
 Nov. 95   & 1995.875 &  25.8 & 49.76 &    31 &   10.6 &  131 &   83 &  188 & 0.028 &   94 \\
 Dec. 95-2 & 1995.962 &  32.7 & 41.47 &    39 &    1.6 &   39 &    0 &  117 & 0.093 &   50 \\
\\
 Jan. 96   & 1996.045 &  29.7 & 49.64 &    34 &    0.0 &    0 &    0 &   42 & 0.095 &   53 \\
 May  96   & 1996.347 &  49.9 & 49.47 &    16 &    4.7 &   70 &   25 &  127 & 0.028 &   98 \\
 Aug. 96   & 1996.615 &  45.0 & 49.26 &    21 &    4.9 &   77 &   31 &  134 & 0.075 &   49 \\
 Oct. 96   & 1996.749 &  45.8 & 49.15 &    21 &    5.9 &   82 &   46 &  127 & 0.053 &   70 \\
 Nov. 96   & 1996.882 &  48.7 & 49.09 &    28 &    1.6 &   22 &    0 &   64 & 0.097 &   45 \\
\\
 Jan. 97   & 1997.019 &  49.8 & 49.04 &    24 &    2.8 &   37 &    6 &   79 & 0.197 &   13 \\
 Mar. 97   & 1997.151 &  44.9 & 48.93 &    23 &    1.6 &   19 &    0 &   55 & 0.457 &    1 \\
 Apr. 97   & 1997.277 &  42.9 & 48.83 &    22 &    3.2 &   41 &   12 &   79 & 0.049 &   79 \\
 June 97   & 1997.403 &  45.6 & 48.78 &    26 &   10.3 &  140 &   91 &  199 & 0.073 &   43 \\
 July 97   & 1997.537 &  45.9 & 48.67 &    22 &    1.6 &   22 &    0 &   56 & 0.445 &    1 \\
 Sep. 97   & 1997.671 &  46.4 & 48.56 &    15 &    3.9 &   62 &   25 &  110 & 0.036 &   91 \\
 Oct. 97   & 1997.803 &  45.0 & 48.45 &    25 &    4.6 &   63 &   28 &  108 & 0.127 &   23 \\
 Dec. 97   & 1997.940 &  47.0 & 48.34 &    22 &    4.7 &   78 &   34 &  135 & 0.054 &   66 \\
\\
 Apr. 98   & 1998.225 &  44.9 & 48.05 &    38 &    5.8 &   82 &   35 &  140 & 0.048 &   77 \\
 May  98   & 1998.347 &  30.0 & 51.17 &    21 &    4.4 &   57 &   24 &   98 & 0.036 &   90 \\
 July 98   & 1998.477 &  45.6 & 51.06 &    21 &    5.7 &   72 &   36 &  118 & 0.076 &   46 \\
 Aug. 98   & 1998.611 &  45.7 & 50.93 &    31 &    4.1 &   52 &   20 &   95 & 0.047 &   82 \\
 Oct. 98   & 1998.745 &  45.8 & 50.81 &    38 &    4.7 &   56 &   18 &  103 & 0.027 &   96 \\
 Nov. 98   & 1998.883 &  45.8 & 50.68 &    30 &    5.2 &   59 &   20 &  107 & 0.078 &   51 \\
\\
 Jan. 99   & 1999.014 &  44.7 & 50.54 &    21 &    2.3 &   29 &    0 &   72 & 0.084 &   51 \\
 Feb. 99   & 1999.130 &  38.7 & 50.43 &    15 &    2.3 &   34 &    4 &   76 & 0.096 &   42 \\
 Apr. 99   & 1999.279 &  51.7 & 50.29 &     9 &    1.5 &   33 &    5 &   74 & 0.054 &   76 \\
 June 99   & 1999.417 &  46.7 & 50.17 &    14 &   14.0 &  185 &  140 &  239 & 0.031 &   98 \\
 July 99   & 1999.551 &  45.7 & 50.06 &    17 &    6.0 &  111 &   54 &  182 & 0.100 &   32 \\
 Sep. 99   & 1999.685 &  45.7 & 49.91 &    20 &    3.3 &   42 &    5 &   93 & 0.250 &    5 \\
 Oct. 99   & 1999.801 &  38.7 & 49.78 &    15 &    9.6 &  134 &   81 &  196 & 0.082 &   49 \\
\\
 Jan. 00   & 2000.035 &  28.8 & 49.59 &    23 &    7.3 &   84 &   46 &  129 & 0.101 &   30 \\
 Feb. 00   & 2000.127 &  30.7 & 49.48 &    20 &    7.9 &   92 &   55 &  138 & 0.044 &   80 \\
 Mar. 00   & 2000.207 &  28.8 & 49.42 &    18 &    9.3 &  106 &   70 &  150 & 0.051 &   72 \\
 May  00   & 2000.359 &  30.7 & 49.24 &    12 &    1.6 &   16 &    0 &   43 & 0.048 &   85 \\
 June 00   & 2000.451 &  33.7 & 49.18 &    16 &    0.8 &   13 &    0 &   59 & 0.324 &    6 \\
 July 00   & 2000.540 &  32.0 & 49.12 &    27 &    6.2 &   66 &   33 &  107 & 0.083 &   42 \\
 Aug. 00   & 2000.626 &  31.3 & 49.06 &    14 &    5.2 &   74 &   41 &  116 & 0.088 &   37 \\
 Sep. 00   & 2000.704 &  27.7 & 49.00 &    30 &    9.0 &  107 &   62 &  160 & 0.091 &   36 \\
 Oct. 00   & 2000.796 &  30.7 & 48.90 &    14 &    0.3 &    4 &    0 &   31 & 0.090 &   56 \\
 Nov. 00   & 2000.876 &  28.7 & 48.84 &    22 &    1.0 &   11 &    0 &   41 & 0.166 &   24 \\
 Dec. 00   & 2000.958 &  30.7 & 48.78 &    25 &    7.4 &   78 &   43 &  119 & 0.066 &   64 \\
\\
 Feb. 01   & 2001.122 &  29.8 & 41.11 &    20 &    6.5 &   80 &   47 &  123 & 0.100 &   29 \\
 Mar. 01   & 2001.214 &  33.4 & 48.53 &    17 &    2.3 &   26 &    0 &   66 & 0.077 &   55 \\
 Apr. 01   & 2001.290 &  22.7 & 48.43 &    16 &    6.7 &   70 &   41 &  107 & 0.087 &   40 \\
 May  01   & 2001.373 &  31.7 & 48.37 &    20 &   12.0 &  118 &   85 &  158 & 0.090 &   35 \\
 June 01   & 2001.469 &  31.7 & 48.27 &    19 &    7.2 &   66 &   38 &   99 & 0.047 &   77 \\
 July 01   & 2001.547 &  23.7 & 48.17 &     7 &    3.0 &   36 &   17 &   65 & 0.026 &   98 \\
 Aug. 01   & 2001.624 &  28.7 & 48.11 &    17 &    7.0 &  117 &   66 &  180 & 0.082 &   41 \\
 Sep. 01   & 2001.701 &  27.7 & 48.06 &    10 &    2.5 &   24 &    4 &   52 & 0.126 &   22 \\
 Oct. 01   & 2001.793 &  30.7 & 47.96 &    12 &    7.0 &   63 &   39 &   94 & 0.120 &   23 \\
 Nov. 01   & 2001.887 &  34.8 & 47.91 &    19 &    4.7 &   39 &   17 &   67 & 0.104 &   29 \\
 Dec. 01   & 2001.955 &  22.8 & 47.86 &    20 &    4.4 &   47 &   22 &   80 & 0.056 &   72 \\
\\
 Jan. 02   & 2002.043 &  29.7 & 47.75 &    31 &   23.2 &  201 &  153 &  254 & 0.162 &   18 \\
 Feb. 02   & 2002.120 &  27.7 & 41.01 &    12 &    7.3 &   78 &   48 &  114 & 0.121 &   24 \\
 Mar. 02   & 2002.199 &  28.8 & 47.62 &    15 &    6.2 &   53 &   28 &   84 & 0.090 &   35 \\
 Apr. 02   & 2002.291 &  30.7 & 47.51 &    13 &    2.7 &   25 &   10 &   46 & 0.127 &   28 \\
 May  02   & 2002.354 &  20.7 & 47.45 &    23 &    6.0 &   63 &   31 &  104 & 0.024 &   99 \\
 June 02   & 2002.448 &  36.8 & 47.40 &    30 &    7.7 &   67 &   39 &  101 & 0.089 &   38 \\
 July 02   & 2002.541 &  29.7 & 47.30 &    16 &    2.0 &   20 &    0 &   50 & 0.070 &   60 \\
 Aug. 02   & 2002.619 &  27.7 & 47.24 &    18 &   12.7 &  126 &   91 &  168 & 0.027 &   97 \\
 Sep. 02   & 2002.698 &  28.7 & 47.18 &    14 &    7.8 &   74 &   48 &  107 & 0.035 &   91 \\
 Oct. 02   & 2002.790 &  30.8 & 47.07 &    16 &    4.7 &   42 &   19 &   70 & 0.030 &   96 \\
 Nov. 02   & 2002.868 &  27.8 & 42.54 &    48 &    6.0 &   61 &   24 &  106 & 0.073 &   59 \\
 Dec. 02   & 2002.947 &  28.8 & 49.58 &    25 &    4.7 &   46 &   19 &   81 & 0.044 &   84 \\
\\
 Jan. 03   & 2003.040 &  30.8 & 49.51 &    15 &    6.9 &   59 &   37 &   86 & 0.106 &   27 \\
 Feb. 03   & 2003.117 &  27.7 & 49.44 &    20 &    5.9 &   53 &   27 &   86 & 0.071 &   50 \\
 Mar. 03   & 2003.199 &  29.8 & 49.38 &    21 &    8.0 &   70 &   44 &  103 & 0.093 &   31 \\
 Apr. 03   & 2003.284 &  27.7 & 49.27 &    22 &    4.7 &   54 &   27 &   89 & 0.122 &   23 \\
 May  03   & 2003.366 &  29.7 & 49.21 &    13 &    7.1 &   66 &   37 &  102 & 0.084 &   40 \\
 June 03   & 2003.448 &  29.7 & 49.16 &    17 &   10.4 &  114 &   77 &  159 & 0.077 &   46 \\
 July 03   & 2003.538 &  29.8 & 49.05 &    21 &   10.1 &  106 &   67 &  154 & 0.068 &   48 \\
 Aug. 03   & 2003.628 &  32.7 & 48.94 &    19 &    2.9 &   32 &    4 &   67 & 0.029 &   96 \\
 Sep. 03   & 2003.713 &  30.7 & 48.94 &    11 &    0.0 &    0 &    0 &   15 & 0.065 &   74 \\
 Oct. 03   & 2003.793 &  24.5 & 48.83 &    20 &   10.0 &  104 &   69 &  147 & 0.057 &   64 \\
 Nov. 03   & 2003.866 &  26.7 & 35.64 &    18 &    3.6 &   47 &   20 &   85 & 0.135 &   23 \\
 Nov. 03-1 & 2003.875 &  26.2 & 13.11 &    10 &    2.4 &   84 &   11 &  187 & 0.066 &   71 \\
 Dec. 03   & 2003.945 &  27.5 & 35.61 &    11 &    5.9 &   78 &   43 &  123 & 0.113 &   27 \\
 Dec. 03-1 & 2003.960 &  26.9 & 13.07 &    10 &    2.1 &   77 &   13 &  170 & 0.023 &   99 \\
\\
 Jan. 04   & 2004.037 &  30.8 & 35.54 &    19 &    0.0 &    0 &    0 &   24 & 0.045 &   89 \\
 Jan. 04-1 & 2004.053 &  17.3 & 13.00 &     7 &    2.4 &  132 &   44 &  278 & 0.077 &   70 \\
 Feb. 04   & 2004.131 &  34.8 & 35.43 &    14 &    5.1 &   62 &   30 &  102 & 0.098 &   41 \\
 Feb. 04-1 & 2004.145 &  18.3 & 13.01 &    10 &    3.5 &  151 &   43 &  287 & 0.091 &   46 \\
 Mar. 04   & 2004.212 &  28.8 & 35.37 &    22 &    4.6 &   59 &   28 &  101 & 0.097 &   39 \\
 Mar. 04-1 & 2004.226 &  30.8 & 12.99 &     8 &    0.0 &    0 &    0 &  163 & 0.200 &   28 \\
 Apr. 04   & 2004.289 &  23.6 & 48.29 &    19 &    4.3 &   67 &   30 &  116 & 0.076 &   41 \\
 May  04   & 2004.354 &  23.4 & 22.03 &     9 &    3.6 &   78 &   25 &  148 & 0.076 &   46 \\
 June 04   & 2004.455 &  38.5 & 22.00 &    10 &    4.1 &   98 &   34 &  182 & 0.091 &   32 \\
 July 04   & 2004.544 &  24.9 & 21.95 &    14 &    2.0 &   43 &    0 &  118 & 0.051 &   74 \\
 Aug. 04   & 2004.623 &  29.3 & 21.93 &    12 &    5.2 &  139 &   80 &  218 & 0.048 &   82 \\
 Sep. 04   & 2004.712 &  32.9 & 42.42 &    14 &    0.1 &    0 &    0 &   25 & 0.103 &   42 \\
 Oct. 04   & 2004.800 &  28.8 & 47.67 &    11 &    1.9 &   22 &    1 &   50 & 0.074 &   56 \\
 Nov. 04-1 & 2004.881 &  29.7 & 47.62 &    17 &    7.3 &   73 &   43 &  111 & 0.037 &   87 \\
 Dec. 04   & 2004.954 &  25.6 & 47.57 &    45 &   25.3 &  305 &  238 &  381 & 0.025 &   96 \\
\\
 Jan. 05   & 2005.047 &  31.7 & 47.47 &    14 &    1.4 &   12 &    0 &   30 & 0.083 &   52 \\
 Feb. 05   & 2005.148 &  21.0 & 47.39 &    11 &    7.8 &   89 &   55 &  131 & 0.242 &   10 \\
 Mar. 05   & 2005.221 &  27.6 & 47.34 &    10 &    3.6 &   35 &   16 &   60 & 0.054 &   70 \\
 Apr. 05   & 2005.283 &  20.6 & 47.29 &    22 &    7.7 &   90 &   52 &  137 & 0.053 &   73 \\
 May  05   & 2005.373 &  31.6 & 47.19 &    18 &    4.9 &   43 &   22 &   69 & 0.079 &   53 \\
 June 05   & 2005.474 &  20.6 & 45.99 &    19 &    1.7 &   19 &    1 &   47 & 0.227 &   13 \\
 July 05   & 2005.545 &  26.7 & 45.93 &    14 &    1.7 &   16 &    3 &   39 & 0.048 &   84 \\
 Aug. 05   & 2005.626 &  24.0 & 45.81 &    19 &    5.0 &   52 &   22 &   91 & 0.118 &   25 \\
 Sep. 05   & 2005.702 &  26.0 & 45.74 &    19 &    4.2 &   40 &   16 &   72 & 0.081 &   47 \\
 Oct. 05   & 2005.781 &  27.0 & 45.67 &    12 &    7.6 &   80 &   49 &  119 & 0.170 &   14 \\
 Nov. 05   & 2005.872 &  30.8 & 45.57 &    22 &   11.7 &  101 &   69 &  140 & 0.045 &   77 \\
 Dec. 05   & 2005.953 &  27.0 & 45.50 &    25 &   12.3 &  106 &   74 &  143 & 0.089 &   32 \\
\\
 Jan. 06   & 2006.046 &  34.8 & 45.45 &    13 &    3.7 &   32 &   15 &   54 & 0.102 &   33 \\
 Feb. 06   & 2006.138 &  29.7 & 45.36 &    30 &    6.9 &   71 &   35 &  114 & 0.059 &   64 \\
 Mar. 06   & 2006.214 &  24.7 & 45.27 &    17 &    2.2 &   20 &    2 &   46 & 0.084 &   47 \\
 Apr. 06   & 2006.281 &  23.7 & 45.22 &    25 &   13.6 &  137 &   98 &  182 & 0.045 &   78 \\
 May  06   & 2006.370 &  33.7 & 45.14 &    16 &    6.4 &   59 &   33 &   92 & 0.043 &   83 \\
 June 06   & 2006.461 &  32.7 & 45.08 &    16 &    5.8 &   56 &   33 &   86 & 0.159 &   16 \\
 July 06   & 2006.546 &  30.7 & 45.06 &    28 &    7.1 &   74 &   34 &  121 & 0.108 &   25 \\
 Aug. 06   & 2006.637 &  29.7 & 44.98 &    21 &    1.6 &   18 &    0 &   54 & 0.129 &   31 \\
 Sep. 06   & 2006.717 &  28.7 & 44.94 &    20 &    8.7 &   91 &   53 &  137 & 0.051 &   71 \\
 Oct. 06   & 2006.796 &  28.7 & 44.89 &    25 &    5.7 &   57 &   31 &   91 & 0.067 &   59 \\
 Nov. 06   & 2006.873 &  23.7 & 50.88 &    30 &   17.0 &  152 &  111 &  199 & 0.056 &   65 \\
 Dec. 06   & 2006.948 &  27.6 & 50.83 &    30 &    6.2 &   69 &   31 &  114 & 0.056 &   70 \\
\\
 Jan. 07   & 2007.043 &  35.6 & 50.77 &    25 &   10.8 &   89 &   57 &  126 & 0.082 &   36 \\
 Feb. 07   & 2007.138 &  30.6 & 50.66 &    25 &    6.8 &   63 &   31 &  103 & 0.093 &   36 \\
 Mar. 07   & 2007.214 &  26.6 & 50.60 &    19 &    4.5 &   41 &   19 &   70 & 0.207 &    7 \\
 Apr. 07   & 2007.279 &  22.7 & 50.55 &    22 &    2.4 &   23 &    3 &   50 & 0.133 &   28 \\
 May  07   & 2007.368 &  30.7 & 50.45 &    19 &    7.4 &   70 &   38 &  108 & 0.069 &   52 \\
 July 07   & 2007.544 &  28.7 & 50.34 &    21 &    3.7 &   36 &   13 &   66 & 0.044 &   83 \\
 Aug. 07   & 2007.637 &  30.8 & 50.24 &    22 &    0.0 &    0 &    0 &   25 & 0.068 &   73 \\
 Sep. 07   & 2007.715 &  27.5 & 50.19 &    24 &   15.0 &  130 &   93 &  172 & 0.131 &   19 \\
 Oct. 07   & 2007.796 &  29.7 & 50.13 &    18 &    4.7 &   37 &   19 &   62 & 0.020 &   99 \\
 Nov. 07   & 2007.886 &  29.7 & 50.02 &    22 &    7.9 &   68 &   42 &  100 & 0.068 &   55 \\
 Dec. 07   & 2007.964 &  26.7 & 49.96 &    20 &    8.8 &   70 &   44 &  101 & 0.043 &   79 \\
\hline
\end{longtable*}

\begin{table*}[t!]
\caption{Approximate cross section for neutrino capture by \nuc{71}{Ga} if
         the contribution of the first two excited states is set to zero.}
\label{sigmavse}
\scriptsize{
\begin{tabular*}{\hsize}{@{} c @{\extracolsep{\fill}} c c c d c c c d c c c @{}}
\hline
\hline 
\mco{$\nu$ energy} & \mc{3}{c}{Cross section $(10^{-46}$ cm$^2)$} & \mco{$\nu$ energy} & \mc{3}{c}{Cross section $(10^{-46}$ cm$^2)$} & \mco{$\nu$ energy} & \mc{3}{c}{Cross section $(10^{-46}$ cm$^2)$} \rule{0pt}{2.5ex} \\ \cline{2-4} \cline{6-8} \cline{10-12}
\mco{(MeV)}        &  Best   & $-1\sigma$& $+1\sigma$             & \mco{(MeV)}        &  Best   & $-1\sigma$& $+1\sigma$             & \mco{(MeV)}        &  Best   & $-1\sigma$  & $+1\sigma$           \\
\hline
   0.240 & 1.310\E{1} & 1.280\E{1} & 1.340\E{1} &   1.445 & 1.944\E{2} & 1.897\E{2} & 2.274\E{2} &   9.500 & 4.749\E{4} & 4.053\E{4} & 6.275\E{4} \rule{0pt}{2.5ex} \\
   0.250 & 1.357\E{1} & 1.326\E{1} & 1.388\E{1} &   1.500 & 2.153\E{2} & 2.104\E{2} & 2.495\E{2} &  10.000 & 5.653\E{4} & 4.820\E{4} & 7.474\E{4} \\
   0.275 & 1.499\E{1} & 1.465\E{1} & 1.533\E{1} &   1.600 & 2.451\E{2} & 2.395\E{2} & 2.859\E{2} &  10.500 & 6.638\E{4} & 5.650\E{4} & 8.785\E{4} \\
   0.300 & 1.662\E{1} & 1.624\E{1} & 1.700\E{1} &   1.700 & 2.771\E{2} & 2.707\E{2} & 3.252\E{2} &  11.000 & 7.703\E{4} & 6.548\E{4} & 1.020\E{5} \\
   0.325 & 1.836\E{1} & 1.794\E{1} & 1.878\E{1} &   1.750 & 2.939\E{2} & 2.871\E{2} & 3.458\E{2} &  11.500 & 8.848\E{4} & 7.510\E{4} & 1.173\E{5} \\
   0.350 & 2.018\E{1} & 1.972\E{1} & 2.064\E{1} &   2.000 & 3.932\E{2} & 3.702\E{2} & 4.712\E{2} &  12.000 & 1.007\E{5} & 8.535\E{4} & 1.336\E{5} \\
   0.375 & 2.208\E{1} & 2.157\E{1} & 2.259\E{1} &   2.500 & 6.428\E{2} & 5.986\E{2} & 7.826\E{2} &  12.500 & 1.137\E{5} & 9.621\E{4} & 1.508\E{5} \\
   0.400 & 2.406\E{1} & 2.351\E{1} & 2.461\E{1} &   3.000 & 9.806\E{2} & 9.043\E{2} & 1.211\E{3} &  13.000 & 1.274\E{5} & 1.077\E{5} & 1.692\E{5} \\
   0.425 & 2.606\E{1} & 2.546\E{1} & 2.799\E{1} &   3.500 & 1.449\E{3} & 1.323\E{3} & 1.811\E{3} &  13.500 & 1.418\E{5} & 1.197\E{5} & 1.884\E{5} \\
   0.450 & 2.810\E{1} & 2.746\E{1} & 3.029\E{1} &   4.000 & 2.108\E{3} & 1.905\E{3} & 2.662\E{3} &  14.000 & 1.569\E{5} & 1.322\E{5} & 2.086\E{5} \\
   0.500 & 3.231\E{1} & 3.157\E{1} & 3.516\E{1} &   4.500 & 3.043\E{3} & 2.722\E{3} & 3.880\E{3} &  14.500 & 1.728\E{5} & 1.454\E{5} & 2.298\E{5} \\
   0.600 & 4.109\E{1} & 4.015\E{1} & 4.585\E{1} &   5.000 & 4.336\E{3} & 3.842\E{3} & 5.573\E{3} &  15.000 & 1.893\E{5} & 1.590\E{5} & 2.520\E{5} \\
   0.700 & 5.027\E{1} & 4.912\E{1} & 5.776\E{1} &   5.500 & 6.072\E{3} & 5.337\E{3} & 7.853\E{3} &  15.500 & 2.064\E{5} & 1.731\E{5} & 2.749\E{5} \\
   0.800 & 6.478\E{1} & 6.329\E{1} & 6.924\E{1} &   6.000 & 8.350\E{3} & 7.290\E{3} & 1.086\E{4} &  16.000 & 2.241\E{5} & 1.875\E{5} & 2.988\E{5} \\
   0.900 & 7.829\E{1} & 7.649\E{1} & 8.470\E{1} &   6.500 & 1.133\E{4} & 9.832\E{3} & 1.479\E{4} &  18.000 & 3.010\E{5} & 2.495\E{5} & 4.025\E{5} \\
   1.000 & 9.299\E{1} & 9.085\E{1} & 1.017\E{2} &   7.000 & 1.515\E{4} & 1.309\E{4} & 1.985\E{4} &  20.000 & 3.860\E{5} & 3.162\E{5} & 5.184\E{5} \\
   1.100 & 1.168\E{2} & 1.142\E{2} & 1.306\E{2} &   7.500 & 1.989\E{4} & 1.712\E{4} & 2.613\E{4} &  22.500 & 5.013\E{5} & 4.028\E{5} & 6.778\E{5} \\
   1.200 & 1.372\E{2} & 1.341\E{2} & 1.549\E{2} &   8.000 & 2.550\E{4} & 2.188\E{4} & 3.357\E{4} &  25.000 & 6.233\E{5} & 4.883\E{5} & 8.501\E{5} \\
   1.300 & 1.593\E{2} & 1.557\E{2} & 1.814\E{2} &   8.500 & 3.198\E{4} & 2.737\E{4} & 4.216\E{4} &  30.000 & 8.701\E{5} & 6.354\E{5} & 1.217\E{6} \\
   1.400 & 1.831\E{2} & 1.789\E{2} & 2.100\E{2} &   9.000 & 3.928\E{4} & 3.357\E{4} & 5.186\E{4} \\
\hline
\hline 
\end{tabular*}
} % end scriptsize
\end{table*}

\section{Modified cross section \break for neutrino capture}
\label{sigma_app}

\vspace*{3.2ex} % used to balance columns

The cross section for neutrino capture by \nuc{71}{Ga} was calculated
by Bahcall \cite{gacross} and is given in his Table~II (best
estimate), Table~III ($3\sigma$ lower limit), and Table~IV ($3\sigma$
upper limit).  Based on information on the contributions of the
excited states given in the text of Bahcall's article, if we assume
the matrix element for neutrino capture to the first two excited
states of \nuc{71}{Ge} to be zero, but that the matrix elements of the
other excited states are unchanged, we can approximate the
best-estimate cross section in various energy regions as follows:
\begin{equation*}
\sigma = \left\{
  \begin{array}{lll}
    13.10 + 91.29(E_{\nu} - 0.24)^{1.157} &\text{for } 0.24  &\!\!\!< E_{\nu} \!< 0.733 \\
    0.946\,      \sigma_{\text{best}}     &\text{for } 0.733 &\!\!\!< E_{\nu} \!< 1.033 \\
    0.953\,      \sigma_{\text{best}}     &\text{for } 1.033 &\!\!\!< E_{\nu} \!< 1.483 \\
    0.96\,\,\,\, \sigma_{\text{best}}     &\text{for } 1.483 &\!\!\!< E_{\nu} \!< 1.983 \\
    0.99\,\,\,\, \sigma_{\text{best}}     &\text{for } 1.983 &\!\!\!< E_{\nu} \!< 30.0
  \end{array}
  \right.
\end{equation*}
\noindent where $\sigma_{\text{best}}$ is the value in Table~II of
Ref.~\cite{gacross}, $\sigma$ is in $10^{-46}$ cm$^2$, and the neutrino
energy $E_{\nu}$ is in MeV.  The results are given in our
Table~\ref{sigmavse}.  The $\pm1\sigma$ limits in Table~\ref{sigmavse}
were obtained in a similar manner from the $3\sigma$ limits given in
Ref.~\cite{gacross}.

%\rule{0pt}{3.6ex} % necessary to separate leading bibliography line from preceding equation

\end{document}